# Data-Driven Measures of High-Frequency Trading

Gbenga Ibikunle [a,b], Ben Moews [a,c], Dmitriy Muravyev [d,e], Khaladdin Rzayev [a,f,g*]

[a]*Edinburgh Centre for Financial Innovations, The University of Edinburgh*

[b]*RoZetta Institute, Sydney*

[c]*Centre for Statistics, The University of Edinburgh*

[d]*Department of Finance, University of Illinois at Urbana-Champaign*

[e]*Canadian Derivatives Institute*

[f]*Koç University*

[g]*Systemic Risk Centre, London School of Economics*

## Abstract

Public data do not identify high-frequency trading (HFT), and standard proxies do not separate liquidity-supplying from liquidity-demanding strategies. We overcome this measurement challenge by training machine learning models on proprietary Nasdaq data to map observed HFT activity to public intraday variables. Applying this mapping, we generate daily measures of liquidity-supplying and liquidity-demanding HFT for all U.S. stocks from 2010 to 2023. The measures largely subsume standard proxies and capture time-series variation that those proxies miss. Using proprietary Euronext Paris data, we provide evidence that the approach generalizes across markets and remains predictive years after training. The 14-year panel lets us study HFT and market quality over time. Supply-side HFT is consistently associated with greater pre-announcement information acquisition, more informed trading, and lower bid-ask spreads, while demand-side HFT is associated with the opposite patterns. During COVID-19, HFT-supplied liquidity remained resilient and its association with lower spreads strengthened.

* Corresponding author. Emails: gbenga.ibikunle@ed.ac.uk (Gbenga Ibikunle), ben.moews@ed.ac.uk (Ben Moews), dmuravy2@illinois.edu (Dmitriy Muravyev), khaladdin.rzayev@ed.ac.uk (Khaladdin Rzayev).

We thank Phil Mackintosh and Heinrich Lutjens at Nasdaq for providing data.

## 1. Introduction

High-frequency trading accounts for a large share of equity trading volume; however, aggregate HFT activity masks strategies with distinct implications for market quality. Many HFT firms act as market makers, using speed to supply liquidity and reduce trading costs (e.g., Hendershott et al. 2011; Menkveld 2013; Brogaard et al. 2015). Other HFT strategies demand liquidity, which can increase adverse selection costs and reduce liquidity (e.g., Easley et al. 2011; Biais et al. 2015; Foucault et al. 2017). The balance between liquidity-supplying and liquidity-demanding HFT can thus shape both trading costs and incentives to acquire and trade on information.

Empirical work has struggled to measure this balance across stocks and over time. Public data do not identify HFT activity, much less distinguish liquidity-supplying from liquidity-demanding HFT. Existing measurement approaches face a tradeoff. Proprietary datasets identify HFT directly, most notably Nasdaq's 120-stock sample from 2008-2009, but cover few stocks over short periods.[1] Public-data proxies, such as the quote-to-trade ratio (e.g., Hendershott et al. 2011), offer broader coverage. However, these proxies combine HFT with algorithmic trading, do not distinguish between the two liquidity roles, and, as we show, mainly capture cross-sectional rather than time-series variation in HFT.

We address this measurement problem by learning how liquidity-supplying and liquidity-demanding HFT appear in public intraday data from a setting in which HFT is directly observed. We train tree-based ensemble models on proprietary Nasdaq data that identify HFT activity, using 24 same-day public variables that measure trading activity, liquidity, and volatility. Thus, instead of imposing a particular public-data proxy as a measure of HFT, the models learn from the data how public intraday variables map into observed HFT activity.

[1] NASDAQ's 120 stock sample from 2008 to 2009 that we use is, arguably, the most popular. Prior studies also use proprietary data from the Investment Industry Regulatory Organization of Canada (IIROC)'s S&P/TSX 60 stocks, and the National Stock Exchange of India (NSE)'s 100-stock dataset from 2015, which are much harder to obtain.

Given that the input public variables can be computed for any TAQ stock-day, we use the trained mapping to construct daily measures of liquidity-supplying and liquidity-demanding HFT for 8,314 U.S. stocks from 2010 to 2023. The two output HFT measures are correlated at 0.50, indicating that they capture related yet distinct HFT activity.

We next test whether our measures add information beyond existing public-data proxies, including the quote-to-trade ratio, odd-lot volume, and trade and quote message counts. We train the models on January-June 2009 data and evaluate them out-of-sample on July-December 2009 data, when HFT activity is still observed. Conventional proxies predict HFT activity when used individually; however, our measures largely subsume them in joint regressions. Moreover, standard proxies mainly rank stocks by average HFT intensity; our measures also track how HFT changes within a stock over time. This time-series variation allows us to study how HFT's association with information acquisition and liquidity changes over time.

We further test whether the measures capture HFT activity rather than simply repackaging public intraday variables. The validation uses market-design changes that shift the incentives for fast trading. NYSE Amex's 2017 speed bump discourages fast trading, while Nasdaq's 2011 data-feed upgrade increases the value of speed; our measures fall after the speed bump and rise after the feed upgrade (Ye et al. 2013; Khapko and Zoican 2021; Aït-Sahalia and Sağlam 2024). The measures also respond to latency-arbitrage opportunities as theory predicts. Aggressive HFT rises as fast traders exploit stale quotes, while passive HFT falls as market makers avoid being picked off (Budish et al. 2015; Foucault et al. 2017).[2]

[2] Additional tests support our measurement approach. The results are similar whether we include quote-level variables or measure HFT activity using unscaled volume rather than HFT share of total volume. The measures also align with theory. Demand-side HFT rises with intermarket sweep volume and falls with market depth, while supply-side HFT rises with depth (Klein 2020; Goldstein et al. 2023). Our tree ensemble approach also outperforms simpler predictive models using the same inputs.

Because our models are trained on 2009 data, a central concern is whether the relationship between public intraday variables and HFT activity remains stable across markets and over time. We examine this concern using proprietary data with HFT identifiers for stocks listed on Euronext Paris. The evidence supports both cross-market and temporal stability. In both markets, HFT activity is predicted mainly by the same public-data signals, especially volume and depth. Within Euronext, these signals remain important in both 2009 and 2017. More directly, a model trained on 2009 Euronext data predicts observed HFT in 2017 as well as it does in 2010, with no deterioration in predictive accuracy over eight years. Together, these tests suggest that the mapping is not unique to Nasdaq or the late 2000s.

We use the 2010-2023 panel to study how the two HFT liquidity roles relate to information acquisition, the mechanisms behind this relationship, and HFT behavior during market-wide stress. Because HFT activity is endogenous to market conditions, we follow Weller's (2018) 2SLS framework, instrumenting each HFT component separately with lagged stock price. Given the identifying assumptions, we emphasize the sign, stability, and relative magnitudes of the estimates rather than interpret them as sharp causal effects.

We first revisit Weller's (2018) finding that algorithmic trading reduces information acquisition before earnings news. His price-jump measure is higher when less information is incorporated into prices before the announcement. Separating HFT by strategy changes the interpretation: a one-standard-deviation increase in HFT demand corresponds to 18% less information acquisition, while the same increase in HFT supply corresponds to 9% more. We show that Weller's (2018) public proxies primarily capture HFT demand, helping reconcile his aggregate result with our findings. Finally, the long panel shows that the pre-2020 decline in information acquisition coincides with rising HFT demand but little growth in HFT supply.

We next examine trading costs as a potential channel for this information acquisition dynamic. Lower trading costs increase the payoff to acquiring and trading on information, while higher costs reduce it. We test this channel in two ways. First, removing the spread from the information-acquisition regressions increases the estimated HFT coefficients, particularly for liquidity-supplying HFT, suggesting that spreads account for part of the relationship. Second, we use quoted spreads as the outcome. A one-standard-deviation increase in liquidity-supplying HFT corresponds to 28% lower spreads, while the same increase in liquidity-demanding HFT corresponds to 38% higher spreads. To account for spread variables appearing in the training feature set, we also reconstruct the HFT measures by excluding them; the spread results remain similar. These opposite relationships are stable over 2010-2023 and are consistent with theories in which fast liquidity suppliers reduce trading costs, while aggressive HFT raises adverse-selection costs (Biais et al. 2015; Budish et al. 2015; Foucault et al. 2016; Baldauf and Mollner 2020; Aït-Sahalia and Sağlam 2024; Jovanovic and Menkveld 2025).

The long panel also lets us study HFT during a sustained market-wide crisis in a mature HFT environment. Prior evidence on HFT under stress comes mainly from short-lived episodes such as the Flash Crash or from 2008, when HFT was still developing. We find that during COVID-19, even as quoted spreads widen sharply and overall market liquidity deteriorates, the association between liquidity-supplying HFT and lower spreads strengthens by 28%, while the corresponding association for liquidity-demanding HFT is unchanged. This evidence suggests that HFT-supplied liquidity is resilient under prolonged stress, consistent with fast traders being better able to provide liquidity when volatility and adverse-selection risk spike.

Finally, we study informed trading using the informed trading intensity measures of Bogousslavsky et al. (2024). We find that a one-standard-deviation increase in HFT demand corresponds to 21% less patient informed trading, while the same increase in HFT supply corresponds to 9% more. The effects for impatient traders are economically small. This concentration among patient traders, who build positions gradually through the limit order book, is consistent with their greater exposure to both aggressive HFT and HFT-provided liquidity. Across these outcomes, separating liquidity supply from demand changes how we interpret aggregate HFT's link to market quality.

The paper contributes to HFT research in three ways. The first contribution is measurement. HFT affects markets through distinct liquidity roles; nevertheless, empirical work has struggled to study these roles at scale. Proprietary datasets identify HFT directly yet cover short periods, narrow cross-sections, or specific venues (Carrion 2013; Hagströmer and Nordén 2013; Menkveld 2013; Brogaard et al. 2014; Nimalendran et al. 2024), and while public-data proxies cover broader samples, they do not separate liquidity supply from liquidity demand (Hendershott et al. 2011; Boehmer et al. 2021a; Chakrabarty et al. 2023).[3] We bridge this gap by constructing separate stock-day measures of liquidity-supplying and liquidity-demanding HFT for U.S. stocks from 2010 to 2023. We make these measures publicly available, providing researchers with a broad panel for studying longer-horizon questions that are difficult to address with existing HFT data. Our approach also extends machine learning methods that scale narrow proprietary data into broader public measures (Easley et al. 2021; Kwan et al. 2021; Bartlett and O'Hara 2024; Bogousslavsky et al. 2024).

Second, the paper shows why separating HFT liquidity supply from demand matters for understanding information production. Prior work documents both benefits and costs of HFT (Biais et al. 2015; Budish et al. 2015; Foucault et al. 2017; Van Kervel and Menkveld 2019; Yang and Zhu 2020; Hirschey 2021); however, aggregate measures combine activities with potentially opposite effects. We show that liquidity-supplying and liquidity-demanding HFT have opposite relationships with information acquisition and informed trading, helping reconcile Weller's (2018) finding that algorithmic trading reduces information acquisition. We also show that these relationships are concentrated among patient informed traders, providing new evidence on how different HFT strategies interact with informed investors.

[3] Related empirical work on HFT, algorithmic trading, and market quality includes Hasbrouck and Saar (2013), Hendershott and Riordan (2013), Chaboud et al. (2014), Brogaard et al. (2015), Conrad et al. (2015), Baron et al. (2019), and Boehmer et al. (2018).

Third, the long panel allows us to study whether the relationship between HFT strategies and market quality persists over time and how HFT-supplied liquidity behaves during periods of sustained market-wide stress. Existing evidence on HFT under stress comes mainly from short-lived events such as the Flash Crash or from 2008, when HFT was still developing (Kirilenko et al. 2017; Brogaard et al. 2018). COVID-19 provides a sustained market-wide crisis in a mature HFT environment. Our evidence suggests that HFT-supplied liquidity remained resilient during this prolonged period of stress, suggesting that fragility documented in short-lived episodes need not generalize to sustained crises.

## 2. Data and Variable Definitions

Our baseline analysis combines two data sources. The first is the Nasdaq HFT dataset used by Brogaard et al. (2014), which identifies the parties to each trade for 120 stocks in 2009 and reports time to the millisecond, price, volume, trade direction, and counterparty type. Trades are classified as HH when both parties are HFTs, HN when an HFT demands liquidity from a non-HFT, NH when a non-HFT demands liquidity from an HFT, and NN when neither party is an HFT. Accordingly, $NASD_HFT_D$ is the share of daily volume in HH and HN trades, and $NASD_HFT_S$ is the share in HH and NH trades; both shares use total HH, HN, NH, and NN volume as the denominator. Thus, HFT demand refers to aggressive executions and HFT supply to passive participation.

The second source is WRDS TAQ Intraday Indicators for the same period. We select 24 public variables covering prices, trading volume, trading costs, liquidity, and volatility, as described in Table 1. Following Bogousslavsky et al. (2024), we use the precomputed WRDS measures to improve replicability and limit researcher discretion. The model uses these TAQ-based inputs to predict the observed HFT shares in the Nasdaq sample; Section 3 describes the estimation.

**INSERT TABLE 1 HERE**

We use several complementary datasets for validation and the subsequent applications. Conventional HFT proxies are constructed from quote-level Millisecond TAQ data. Intraday transactions and corresponding bid and ask quotes come from LSEG Datascope Tick History. Earnings and merger-and-acquisition announcement dates come from I/B/E/S and the Thomson Reuters Securities Data Company database, respectively. Stock returns and trading volume come from CRSP. Finally, proprietary Eurofidai-Bedofih data identify HFT activity in Euronext Paris stocks; Section 5 describes that sample.

**INSERT TABLE 2 HERE**

Table 2 reports definitions and summary statistics for the observed Nasdaq HFT shares, the ML-generated measures, and the variables used in the empirical analyses. $NASD_HFT_D$ and $NASD_HFT_S$ average 0.331 and 0.250, respectively, and the difference is significant at the 1% level. Their distributions are skewed in opposite directions, a pattern also present in the ML-generated $HFT_D$ and $HFT_S$ measures. The relative bid-ask spread averages 0.142% and ranges up to 0.885%, indicating substantial variation in liquidity conditions. The long U.S. sample covers 8,314 stocks in TAQ.

## 3. Methodology

This section explains how we map public intraday data into measures of liquidity-demanding and liquidity-supplying HFT. The model learns that mapping in the 2009 Nasdaq sample, where HFT activity is directly observed, and we then apply the estimated mapping to stock-days for which the same public inputs are available.

### *3.1. Ensemble methods for HFT prediction*

We start by explaining the prediction setup. For each stock-day in the Nasdaq data, the outcomes are the observed shares of volume attributable to liquidity-demanding HFT and liquidity-supplying HFT. The predictors are the 24 public intraday variables in Table 1. No economic sign or functional form is imposed on the relation between an input and HFT activity.

We use random forests and extremely randomized trees (Ho 1995; Geurts et al. 2006). A single tree partitions stock-days using simple rules based on the input variables and assigns each resulting group a predicted HFT share. Trees are useful here because the same market characteristic may be informative only above a threshold or in combination with another characteristic. Averaging predictions across many randomized trees captures these nonlinearities and interactions while reducing estimation noise. Both models minimize mean squared prediction error.

We assess performance out of sample. For model tuning, each run draws 10,000 stock-days, assigns 75% to estimation and 25% to testing, and evaluates predictions only in the held-out observations. We repeat this random split ten times and compare average out-of-sample $R^2$ and its dispersion across runs. This procedure selects a model that predicts observations not used to fit it, rather than one that merely describes the training sample.

We select the ensemble size and minimum number of observations required for a split using this held-out performance. Online Appendix Table OA.A.1 reports the parameter grid and shows that the leading specifications use larger ensembles and fine terminal partitions. Our main temporal out-of-sample test is stricter: we train the selected specification on January-June 2009 and predict observed HFT activity in July-December 2009. Section 3.2 reports those predictions against conventional HFT proxies.

Extremely randomized trees provide the highest out-of-sample accuracy with low dispersion; therefore, we use them for the main measures. Online Appendix OA.A.2 reports the comparison with LASSO, support vector machines, neural networks, and alternative single-and multi-target designs. After selecting the specification, we optimize and fit the mapping in the labeled Nasdaq data and apply it to the same public inputs for the broader U.S. sample.

### *3.2. Comparison with common HFT proxies*

We compare the ML-generated measures with five conventional public-data proxies. $Flick$ is quote volatility, calculated as the stock-day average of the standard deviation of quote midpoints over 100-millisecond intervals (Hasbrouck 2018). $OLV$ is the daily volume of trades smaller than 100 shares (Weller 2018). $QuoteInt$ counts changes in the best quotes or displayed depth (Conrad et al. 2015). $QT$ is the ratio of quoted shares to traded shares (Hendershott et al. 2011; Weller 2018), and $MG$ is the daily number of trade and quote messages (Hendershott et al. 2011; Boehmer et al. 2018). We construct all five from Millisecond TAQ.

To keep the comparison out of sample, we train the ML model on January-June 2009, hold the fitted mapping fixed, and evaluate its predictions in July-December 2009. For the latter period, we regress the observed Nasdaq liquidity-demanding and -supplying HFT shares on the ML prediction and the five common proxies, with stock and day fixed effects:

$$NASD_HFT_D_{i,t} = \alpha_i + \beta_t + \gamma_1 HFT_D_{i,t} + \gamma_2 Flick_{i,t} + \gamma_3 OLV_{i,t} + \gamma_4 QuoteInt_{i,t} + $$
$$+\gamma_5 QT_{i,t} + \gamma_6 MG_{i,t} + \varepsilon_{i,t} \qquad (1)$$

$$NASD_HFT_S_{i,t} = \alpha_i + \beta_t + \gamma_1 HFT_S_{i,t} + \gamma_2 Flick_{i,t} + \gamma_3 OLV_{i,t} + \gamma_4 QuoteInt_{i,t} + $$
$$+\gamma_5 QT_{i,t} + \gamma_6 MG_{i,t} + \varepsilon_{i,t} \qquad (2),$$

We first estimate a separate regression for each candidate measure and then include all measures jointly as in Equations (1) and (2). The regressors are standardized, while the dependent variables are not. We evaluate the horse race using $t$-statistics, within-$R^2$, and the incremental explanatory power each measure contributes in the joint specification. Standard errors are double-clustered by stock and date.

**INSERT TABLE 3 HERE**

Panel A of Table 3 reports the liquidity-supplying results with stock and day fixed effects. $HFT_S$ has the largest $t$-statistic and produces the highest within-$R^2$ among the candidate measures. The flickering-quote and odd-lot proxies are not statistically significant. In the joint regression, $HFT_S$ remains precisely estimated, and adding the five conventional proxies increases within-$R^2$ only slightly. Thus, the conventional measures contribute little explanatory power beyond $HFT_S$.

Panel B reports the corresponding liquidity-demanding results. Prediction is harder on the demand side, as indicated by lower within-$R^2$ values, whereas $HFT_D$'s coefficient estimate, again, has the largest $t$-statistic and the highest within-$R^2$. Among the conventional proxies, only the quote-to-trade ratio enters positively in the fixed-effects regressions. In the joint specification, $HFT_D$ retains most of its explanatory power.

The stock and day fixed effects in Panels A and B remove persistent differences across stocks and common daily shocks. The remaining association therefore tests whether a measure tracks changes in observed HFT within a stock over time. The ML-generated measures perform strongly, whereas the conventional proxies have weaker relationships with observed HFT.

Panels C and D retain day fixed effects but omit stock fixed effects, allowing persistent cross-sectional differences to contribute to the estimates. Quote intensity, the quote-to-trade ratio, and message count then become substantially more informative for both HFT components. This contrast indicates that conventional proxies are better at ranking stocks by average HFT intensity than at tracking within-stock changes. Even in this specification, the ML-generated measures have the largest $t$-statistics and within-$R^2$ values and largely subsume the conventional proxies in the joint regressions.

Overall, the out-of-sample horse race favors the ML-generated measures on the criteria that are comparable across proxies: statistical precision, within-$R^2$, and incremental explanatory power. Although the conventional measures retain useful cross-sectional information, the ML predictions better capture both cross-sectional differences and time-series variation in liquidity-demanding and liquidity-supplying HFT.

### *3.3.Model assessment and extrapolation to U.S. stocks*

We apply the estimated model in the same way that fitted coefficients from an asset-pricing model are applied to new returns. For each stock-day, we supply the 24 public variables in Table 1 and obtain predicted $HFT_D$ and $HFT_S$. This step assumes that the Nasdaq mapping remains informative for other U.S. stocks and later dates. Section 4 evaluates this assumption within the U.S. market. Section 5 examines the same broader concern using independently labeled Euronext data, testing whether the relationship between public intraday variables and HFT activity is stable across markets and over time. The resulting U.S. panel contains 9,396,251 stock-days from January 4, 2010 through October 18, 2023.

Figures 1 and 2 provide an economic reading of the fitted model. Feature importance measures how much a variable contributes to prediction within the tree ensemble; it is not a test of statistical significance or causality. Partial dependence plots summarize how the fitted prediction changes with one input while averaging over the observed values of the other inputs. We therefore use both figures to interpret the signals learned by the model, not to estimate the effect of a market characteristic on HFT.

**INSERT FIGURE 1 HERE**

Figure 1 shows a concentrated predictive hierarchy. Trading volume, market depth, and intermarket sweep order (ISO) activity contribute most to the predictions. Volume identifies the overall intensity of HFT: trade counts and dollar volume identify stock-days with enough activity for rapid intermediation and repeated execution. Depth helps the model distinguish the two liquidity roles, while ISO activity is especially informative about aggressive HFT.

Spreads, volatility, returns, and institutional order imbalance contribute less incremental predictive information. This ranking does not imply that these variables are economically unimportant. Rather, trade counts, depth, and ISO activity appear to leave a more distinctive footprint of HFT after the other inputs are considered, whereas spreads, volatility, and returns reflect the activity of many market participants.

**INSERT FIGURE 2 HERE**

Figure 2 clarifies the economic content of the leading signals. Predicted $HFT_D$ and $HFT_S$ both rise with the total number of trades, but at a decreasing rate. Volume therefore identifies broad HFT intensity rather than one side of HFT alone, consistent with HFT being more prevalent in actively traded stocks (Brogaard et al. 2014).

ISO activity separates the strategies more sharply. Predicted $HFT_D$ rises strongly with ISO dollar volume, whereas predicted $HFT_S$ changes little. ISOs prioritize immediate execution across venues; hence, this pattern is consistent with aggressive fast trading, including trading against stale quotes and responding rapidly to news (Klein 2020; Li et al. 2021b).[4] It need not imply that every ISO is submitted by an HFT. The model simply uses ISO activity as a signal of market conditions in which aggressive HFT is more prevalent.

Depth provides the complementary distinction. Predicted $HFT_S$ rises as the order book deepens, whereas predicted $HFT_D$ falls. The pattern is economically intuitive: passive HFT contributes orders to the book and is more prevalent in deep markets, while aggressive HFT consumes displayed liquidity and is more prevalent when the book is thinner. It also matches the direct evidence in Goldstein et al. (2023).

The central lesson is simple. Volume identifies overall HFT intensity; depth helps distinguish liquidity supply from liquidity demand; and ISO activity identifies aggressive HFT. The curved relationships also explain why a flexible prediction rule is useful here. A linear specification could approximate average slopes; however, it would not automatically accommodate the thresholds, flattening, and interactions evident in Figure 2.

[4] https://tabbforum.com/opinions/why-hfts-have-an-advantage-part-3-intermarket-sweep-orders/

Online Appendix OA.B.1 reports complementary within-stock correlations. Both measures are positively correlated with market capitalization and dollar volume. Depth provides the clearest difference: its correlation is 0.356 with $HFT_S$ and -0.019 with $HFT_D$, consistent with the interpretation above. $HFT_D$ and $HFT_S$ are correlated at 0.497, so they share a common activity component but retain economically distinct variation. We next assess whether that variation behaves as HFT should in validation tests.

## 4. Empirical Validation

### *4.1.HFT around exchange technology changes.*

We first test whether $HFT_D$ and $HFT_S$ move in the predicted directions around two exchange technology changes: a Nasdaq data-feed upgrade near the training period and an Amex speed bump several years later. These settings provide indirect validation because each change alters the value of trading speed. A measure that tracks HFT should rise when latency falls and decline when a deliberate delay is introduced.

Nasdaq reduced the dissemination latency of its trading data from three milliseconds to one millisecond in October 2011 (Ye et al. 2013). The change was staggered: stocks with tickers beginning with A or B were upgraded on October 10, and the remaining Nasdaq stocks on October 17. We use this implementation schedule to estimate the following stock-day regressions:

$$HFT_D_{i,t} = \alpha_i + \beta_t + \gamma_1 Post_{i,t} + \sum_{k=1}^{4} \delta_{i,t}^{k} C_{i,t}^{k} + \varepsilon_{i,t} \quad (3)$$

$$HFT_S_{i,t} = \alpha_i + \beta_t + \gamma_2 Post_{i,t} + \sum_{k=1}^{4} \delta_{i,t}^{k} C_{i,t}^{k} + \varepsilon_{i,t} \quad (4),$$

where $HFT_D$ and $HFT_S$ are the liquidity-demanding and liquidity-supplying measures. Stock and day fixed effects absorb persistent stock characteristics and common daily shocks. $Post$ equals one after the applicable upgrade date for each Nasdaq-listed stock and remains zero for NYSE- and Amex-listed control stocks. Following Ye et al. (2013), we use a 10-trading-day window around the implementation dates and double-cluster standard errors by stock and day. The controls are daily transaction-return volatility, relative quoted spread, inverse price, and dollar trading volume, all from TAQ.

The second setting is the 700-microsecond round-trip delay implemented by Amex on July 24, 2017, after SEC approval. The delay applies to both inbound and outbound communication and reduces the advantage of speed on the exchange. If the measures track fast trading, both should decline for Amex-listed stocks relative to NYSE- and Nasdaq-listed controls. We estimate:

$$HFT_D_{i,t} = \alpha_i + \beta_t + \gamma_1 Post_{i,t} * Amex_{i,t} + \sum_{k=1}^{4} \delta_{i,t}^{k} C_{i,t}^{k} + \varepsilon_{i,t} \quad (5)$$

$$HFT_S_{i,t} = \alpha_i + \beta_t + \gamma_2 Post_{i,t} * Amex_{i,t} + \sum_{k=1}^{4} \delta_{i,t}^{k} C_{i,t}^{k} + \varepsilon_{i,t} \quad (6),$$

where $Post$ equals one on and after July 24, 2017, and $Amex$ identifies Amex-listed stocks. Their interaction is the treatment variable; the main effects are absorbed by the day and stock fixed effects. We again use a 10-trading-day window, the controls defined above, and standard errors double-clustered by stock and day.

As the HFT measures aggregate trading across venues, substitution away from a treated exchange would attenuate the estimated response toward zero.

**INSERT TABLE 4 HERE**

Table 4 reports the results. Columns (i) and (ii) show that both HFT measures rise after Nasdaq's latency reduction, while columns (iii) and (iv) show that both fall after the Amex's speed bump, relative to stocks listed on the control exchanges. All four estimates have the predicted signs and are statistically significant.

The economic magnitudes are larger for the Amex speed bump, which directly delays exchange communication, than for the Nasdaq upgrade, which accelerates only the consolidated feed while faster direct feeds remain available. Relative to their pre-event averages, $HFT_D$ and $HFT_S$ fall by 2.8% and 4.6% after the speed bump. The corresponding increases after the Nasdaq upgrade are 0.7% and 1.1%. This ordering is consistent with the difference in how directly the two changes affect the value of speed.

These results provide three validation insights. First, the measures respond in the predicted directions to technology changes both near the 2009 training period and in 2017. The evidence therefore supports temporal stability in the U.S. application. Section 5 provides a more direct test using observed HFT on Euronext. Second, the measures distinguish changes that increase the value of speed from a change that reduces it. This sign pattern is useful validation of the measures, although the event estimates do not by themselves identify the broader effects of HFT on market quality.

Third, the speed-bump results complement Aït-Sahalia and Sağlam (2024), who document wider quoted spreads and lower liquidity after the delay. Both HFT components decline in our estimates, with the larger reduction on the supply side. This pattern is consistent with their market-making channel and supports the economic interpretation of $HFT_D$ and $HFT_S$. The next test uses latency-arbitrage opportunities to sharpen that distinction.

### *4.2. HFT and latency arbitrage opportunities.*

Latency arbitrage allows fast traders to act on public information and execute against quotes that have become stale before slower traders can update them (Budish et al. 2015; Foucault et al. 2017; Shkilko and Sokolov 2020; Aquilina et al. 2022). These opportunities should attract aggressive HFT, while the resulting adverse-selection risk may induce fast market makers to reduce liquidity supply (Foucault et al. 2017; Menkveld and Zoican 2017; Baldauf and Mollner 2020). We therefore expect $HFT_D$ to rise and $HFT_S$ to fall as latency-arbitrage opportunities become more frequent.

We test these predictions with the following stock-day regressions:

$$HFT_D_{i,t} = \alpha_i + \beta_t + \gamma_1 NLAO_{i,t} + \sum_{k=1}^{4} \delta_{i,t}^{k} C_{i,t}^{k} + \varepsilon_{i,t} \quad (7)$$

$$HFT_S_{i,t} = \alpha_i + \beta_t + \gamma_2 NLAO_{i,t} + \sum_{k=1}^{4} \delta_{i,t}^{k} C_{i,t}^{k} + \varepsilon_{i,t} \quad (8),$$

where $NLAO$ is the number of latency-arbitrage opportunities. Following Budish et al. (2015), we compare the current midpoint with the previous best quotes to detect latency arbitrage. With $TickSize$ set to \$0.01, an upward opportunity occurs when $Midprice(\tau) > Ask(\tau - 1) + TickSize$, and a downward opportunity occurs when $Midprice(\tau) < Bid(\tau - 1) - TickSize$. In the first case, a fast trader can submit a marketable limit buy order; in the second, a marketable limit sell order. We identify these events from first-level LSEG Datascope quotes. Because processing the full 8,314-stock universe is computationally prohibitive, we use the 120 stocks in the original Nasdaq HFT sample from 2010 through 2023. As Table 2 reports, $NLAO$ averages 68 per stock-day, with a standard deviation of 169 and a maximum of 1,211.

**INSERT TABLE 5 HERE**

Table 5 reports a positive association between $NLAO$ and $HFT_D$, significant at the 1% level, and a negative association between $NLAO$ and $HFT_S$, significant at the 5% level. A one-standard-deviation increase in $NLAO$ is associated with a 1.0% increase in $HFT_D$ and a 1.6% decrease in $HFT_S$ relative to their sample means. The opposing signs match the predicted responses of aggressive fast traders and fast liquidity suppliers, providing additional validation that the two measures capture economically distinct HFT strategies.

## 5. Cross-Market and Temporal Validation

The usefulness of our approach depends on whether the relationship between public intraday variables and HFT activity is stable across markets and over time. Section 4 provides indirect evidence from U.S. market-structure shocks, both near and years after the training period.[5] We now use data from the Euronext Paris exchange provided by the Eurofidai-Bedofih

[5] Prior literature also finds that results based on the Nasdaq HFT dataset for the late 2000s typically hold for other US exchanges (Shkilko and Sokolov 2020), in the post-2009 period (Aït-Sahalia and Sağlam 2024; Brogaard et al. 2025), and internationally (Benos and Sagade 2016; Malceniece et al. 2019; Chakrabarty et al. 2026). Indeed, core HFT strategies have remained stable despite technological advances (Brogaard et al. 2014; Malceniece et al. 2019). The features of HFT strategies developed in recent theories (Li et al. 2021a) are similar to those from a

to provide direct evidence of the temporal stability of our method and assess whether it generalizes to a different market setting, where, as in our Nasdaq training data, HFT activity is observed directly at the trader level.

The Eurofidai-Bedofih database[6] identifies the HFT status of each buyer and seller, the trade initiator, and whether an HFT participant is classified by the Autorité des Marchés Financiers as Pure HFT or Mixed HFT. The HFT flag is available through 2017.[7] The regulatory classification is based on the speed and frequency of order cancellations.[8] These data are well suited to our validation because they provide directly observed liquidity-demanding and liquidity-supplying HFT shares in a major market outside the United States. Their 2009-2017 coverage also allows us to examine stability over an eight-year period. To our knowledge, they are among the most recent commercially available data for a major equity market that identify the HFT status of both trade counterparties and the trade initiator.

We use 2009, 2010, and 2017. The first is the earliest year with the AMF HFT flag and matches the Nasdaq training period, 2010 provides the nearest out-of-sample year, and 2017 is the last year with the AMF HFT flag. The sample covers CAC All-Tradable constituents. For each stock-day, $HFT_D$ is the share of continuous-trading volume for which an HFT initiates the trade, and $HFT_S$ is the share for which an HFT is passive. We exclude pre-arranged crosses because they have no identified initiator.

We construct the public input variables from LSEG Datascope rather than Eurofidai-Bedofih, following the same approach of constructing input features from non-proprietary data as in the U.S. analysis. Two inputs require adjustment. Euronext has no ISO designation; thus, decade ago (Biais et al. 2015; Foucault et al. 2017) and many recent studies continue to rely on datasets from 2009-2012 (Boehmer et al. 2018; Goldstein et al. 2023; Nimalendran et al. 2024).

[6] Recent papers using this data include Colliard and Hoffmann (2017), Bernales et al. (2024), Bellia et al. (2025), and Daures et al. (2026).

[7] The underlying trade data is still available after 2017, but the HFT flag is not included.

[8] A participant receives the HFT designation if the average lifetime of its cancelled orders falls below the market-wide average with at least 100,000 cancellations per year, or if it cancels at least 500,000 orders with lifetimes below 0.1 seconds and the top percentile of lifetimes is below 500 microseconds. Like the Nasdaq classification, these criteria identify HFT participants from observable characteristics of their trading behaviour.

we use the euro value of aggressive trades executed through the quote as a proxy for orders that prioritize immediacy over price improvement. Retail volume is unavailable, but it has low importance in the Nasdaq model; we therefore omit it from both markets when comparing feature importance.

The merged sample contains 490 firms. We estimate extremely randomized trees as in Section 3. The evidence has two distinct parts. First, models trained separately on 2009 Nasdaq and 2009 Euronext data reveal whether similar public signals identify HFT across markets. Second, a model trained on 2009 Euronext data is evaluated against observed Euronext HFT in 2010 and 2017, providing a direct within-market test of temporal stability.

**INSERT FIGURE 3 HERE**

Figure 3 reports the cross-market comparison. In both 2009 models, volume and depth dominate feature importance, while spreads, order imbalance, price and returns, and volatility contribute less. Within the volume group, the number of trades ranks first, followed by local-currency volume and institutional volume in both markets. The leading depth variables differ, which is unsurprising given differences between the Euronext and Nasdaq order books.

Panel B is consistent with Panel A: after averaging importance within economic categories, volume and depth lead in both markets, suggesting that the signals used by the Nasdaq mapping are also relevant for identifying HFT on Euronext.

**INSERT FIGURE 4 HERE**

Figure 4 turns to stability within Euronext. Models trained separately in 2009 and 2017 continue to place volume and depth first, and the ranking of the broader feature groups is stable. This comparison shows that the public signals associated with observed Euronext HFT change little over eight years, although feature rankings alone do not establish predictive accuracy out of sample.

**INSERT TABLE 6 HERE**

Table 6 provides that direct test. We train the model on 2009 Euronext observations and use it, without re-estimation, to predict observed HFT activity in 2010 and 2017. We then regress actual Euronext HFT on the predictions separately by year, including stock and day fixed effects and double-clustering standard errors by stock and day. The comparison asks whether prediction deteriorates as the evaluation date moves farther from 2009.

The predicted measures are positively associated with actual HFT in both years, and all four coefficients are significant at the 1% level. For $HFT_S$, the coefficients are 0.041 in 2010 and 0.045 in 2017 and are not statistically different. For $HFT_D$, the coefficient rises from 0.053 to 0.077; the difference is marginally significant at 10%. Within-$R^2$ also does not decline: it rises from 0.07% to 0.12% for supply and from 0.30% to 0.34% for demand. These within-$R^2$ values are small because stock and day fixed effects remove most cross-sectional and aggregate time-series variation. The corresponding out-of-sample training $R^2$, which uses total variation, is 72% on Euronext and 82% on NASDAQ.

Two conclusions follow from these exercises. Similar signals identify HFT in 2009 Nasdaq and Euronext data, providing cross-market evidence on the economic content of the inputs. More directly, a model trained on 2009 Euronext data remains predictive of observed Euronext HFT in 2017. Together with the U.S. event tests in Section 4, the results reduce the concern that the mapping reflects only idiosyncrasies of one market or a short training period.

## 6. Applications

We use the long U.S. panel to study three related market quality outcomes: information acquisition before public news, bid-ask spreads, and informed trading. Separating HFT supply from demand is central because the two activities can create opposing incentives for investors who acquire and trade on information. The panel also allows for annual stability tests and an examination of liquidity provision during COVID-19, settings beyond the reach of short proprietary samples.

*6.1.HFT and information acquisition*

Price discovery includes both incorporating existing news into prices and producing or acquiring new fundamental information (O'Hara 2003; Brunnermeier 2005; Weller 2018; Brogaard and Pan 2022). Most HFT research examines the first margin and generally finds that fast trading accelerates the incorporation of available information (Menkveld 2016). We focus on the second: whether investors acquire information before a public announcement.

This outcome is measured at low frequency and therefore requires many firms and years. Proprietary data that distinguish aggressive from passive HFT are typically too narrow for this purpose. For example, the Nasdaq sample yields only 480 firm-quarter earnings events. Our measures provide the needed coverage while retaining the supply-demand distinction.

The distinction matters because theory predicts opposing relationships. Liquidity-supplying HFT can lower execution costs and raise the expected return to acquiring information (Menkveld 2013; Brogaard et al. 2015; Aït-Sahalia and Sağlam 2024). Liquidity-demanding HFT can instead anticipate or react to informed order flow through back-running and latency arbitrage strategies, increasing informed investors' trading costs and weakening their incentive to gather information (Van Kervel and Menkveld 2019; Yang and Zhu 2020; Hirschey 2021).

Weller (2018) measures pre-announcement information acquisition with the price jump ratio, $JUMP$: the announcement-window abnormal return divided by the abnormal return accumulated over the pre-announcement and announcement windows. A higher ratio means that more of the price adjustment occurs at public release and therefore indicates less information acquisition beforehand. Using aggregate algorithmic-trading proxies, Weller (2018) finds a positive relationship between algorithmic trading and $JUMP$.

Aggregate proxies do not distinguish liquidity supply from demand; therefore, they cannot reveal whether Weller's (2018) result combines opposing channels. Our contribution is to estimate the two relationships separately and then examine which HFT component is most closely related to Weller's (2018) public proxies.

HFT activity is endogenous to the liquidity and trading characteristics that may also shape information acquisition. For comparability, we follow Weller's (2018) 2SLS design and instrument each HFT component with lagged log stock price, controlling for market capitalization, volatility, effective spread, and institutional order imbalance. The exclusion restriction is not directly testable: conditional on these controls and month fixed effects, lagged price must affect $JUMP$ only through the instrumented HFT component. We therefore present the estimates as IV evidence under this assumption and emphasize their signs, stability, and relative magnitudes rather than interpret them as structural causal effects. The single instrument also requires us to estimate $HFT_D$ and $HFT_S$ in separate equations:

$$HFT_D_{i,m} = \beta_m + \gamma_1 LnPrice_{i,m-1} + \sum_{k=1}^{4} \delta_{i,m-1}^{k} C_{i,m-1}^{k} + \varepsilon_{i,m} \quad (9)$$

$$HFT_S_{i,m} = \beta_m + \gamma_1 LnPrice_{i,m-1} + \sum_{k=1}^{4} \delta_{i,m-1}^{k} C_{i,m-1}^{k} + \varepsilon_{i,m} \quad (10)$$

$$JUMP_{i,m} = \beta_m + \gamma_1 \widehat{HFT_D}_{i,m} + \sum_{k=1}^{4} \delta_{i,m-1}^{k} C_{i,m-1}^{k} + \varepsilon_{i,m} \quad (11)$$

$$JUMP_{i,m} = \beta_m + \gamma_1 \widehat{HFT_S}_{i,m} + \sum_{k=1}^{4} \delta_{i,m-1}^{k} C_{i,m-1}^{k} + \varepsilon_{i,m} \quad (12),$$

where $JUMP$ is the ratio of cumulative abnormal returns over [-1, 1] to cumulative abnormal returns over [-21, 1] around quarterly earnings announcements. $HFT_D$ and $HFT_S$ are averages over [-21, -1]. Lagged price and the controls – volatility, relative effective spread, market value, and institutional order imbalance for trades above \$20,000 – are averaged over [-42, -22]. Following Weller (2018), the specifications include month but not stock fixed effects because stock fixed effects would absorb most variation in the persistent price instrument. Variables are scaled by their root mean square about zero without demeaning.

**INSERT TABLE 7 HERE**

Columns (1) and (2) of Table 7 report the first stages. Lagged price can relate to HFT in either direction. A one-cent tick is smaller relative to price for high-priced stocks, so quotes become stale after smaller fundamental-value changes and may attract fast liquidity demand. For low-priced stocks, the same tick is large relative to price and can create rents from queue priority, potentially attracting fast liquidity supply (Yao and Ye 2018).

The estimates have the predicted opposing signs: log price is positively related to $HFT_D$ (0.221, $t$ = 28.17) and negatively related to $HFT_S$ (-0.453, $t$ = -22.97). The first-stage $F$-statistics are 793.49 and 527.80, respectively, well above conventional weak-instrument thresholds. These patterns are consistent with stale-quote trading on the demand side and tick-size rents on the supply side; they do not, by themselves, establish that these are the only mechanisms generating the first stages.

The positive loading of $HFT_D$ on price also helps interpret Weller's (2018) first stage. His algorithmic-trading proxies load positively on price, in the same direction as our liquidity-demanding measure. This is initial, not dispositive, evidence that his aggregate proxies place more weight on aggressive HFT activity.

Columns (3) and (4) report the second stages. Instrumented $HFT_D$ is positively associated with $JUMP$ (0.469, $t$ = 6.97), whereas instrumented $HFT_S$ is negatively associated with $JUMP$ (-0.229, $t$ = -6.89). A one-standard-deviation increase in the corresponding first-stage fitted value is associated with an 18% increase in $JUMP$ for $HFT_D$ and a 9% decrease for $HFT_S$, relative to the sample mean.[9]

The opposite signs are consistent with the proposed trading-cost channel: aggressive HFT may raise the cost of trading on acquired information, while passive HFT may lower it. The estimates therefore show why combining the two activities can obscure economically distinct relationships. Given the identifying assumption, however, they should be read as supporting the mechanism rather than as sharp causal estimates of each strategy's effect.

[9] Economic magnitudes are computed by multiplying the second-stage coefficient by one standard deviation of the fitted values from the corresponding first stage and expressing the result relative to the sample mean of the dependent variable. As all variables are scaled by a constant, the resulting percentages are identical to those obtained from unscaled data.

Due to $HFT_D$ and $HFT_S$ being instrumented in separate regressions, their coefficients cannot be mechanically added to recover an aggregate effect. Online Appendix E uses the heteroskedasticity-based instruments of Lewbel (2012) to include both components jointly. The coefficients retain their opposing signs, and the sum is positive for $JUMP$. We view this as a useful robustness check suggesting that the demand-side relationship is larger in this sample, not as a structural estimate of the net causal effect of HFT.

This decomposition offers a possible reconciliation with Weller (2018) as his positive association between algorithmic trading and $JUMP$ may arise because the public proxies used in that study are more closely related to liquidity-demanding than liquidity-supplying HFT. The shared positive price first stage points in that direction. Table 8 examines the link directly.

**INSERT TABLE 8 HERE**

Table 8 regresses Weller's (2018) cancel-to-trade ratio, odd-lot rate, and trade-to-order ratio on our two measures.[10] The cancel-to-trade ratio and odd-lot rate are positively related to $HFT_D$ and negatively related to $HFT_S$; the trade-to-order ratio, an inverse algorithmic-trading proxy, has the opposite pattern. Together with Weller's (2018) reported signs, these results indicate that his proxies predominantly reflect liquidity-demanding HFT in our sample. They support, while not uniquely establishing, the demand-side interpretation of his aggregate result.

We next use the longer panel to examine the time-series pattern documented by Weller (2018). His sample shows declining pre-announcement information acquisition through 2016. Our data extend the comparison through 2022.

---

[10] $CT_{i,q}$ is the ratio of cancelled messages to trade messages, $OLR_{i,q}$ measures the proportion of trades below 100 shares, and $TO_{i,q}$ is calculated as the ratio of executed shares to submitted shares.

Panel A of Figure 5 plots annual means and medians of $JUMP$ from 2010 to 2022.[11] Both rise from about 0.42 early in the sample to roughly 0.60 in 2019, consistent with less information being incorporated before earnings announcements. $JUMP$ then falls sharply in 2020 and partially rebounds in 2021-2022.

**INSERT FIGURE 5 HERE**

Changes in disclosure, analyst coverage, passive ownership, and other market conditions may also contribute to this trend. We therefore examine whether changes in HFT activity are associated with the trend, without attributing the aggregate change to HFT alone. We first re-estimate the 2SLS specifications by year and then compare annual HFT activity with $JUMP$.

Panel B of Figure 5 shows stable signs in the annual IV estimates: $HFT_D$ is positively associated with $JUMP$ in every year from 2010 to 2022, while $HFT_S$ is negatively associated with $JUMP$ in every year. Magnitudes vary within a fairly narrow range, and most confidence intervals exclude zero. Thus, the opposing IV relationships are persistent, though the annual estimates do not imply a time-invariant structural effect.

Figure 6 provides complementary level evidence. From 2010 to 2019, average $HFT_D$ rises from roughly 0.27 to 0.31 alongside $JUMP$, while $HFT_S$ remains near 0.19. In 2020-2021, $HFT_S$ rises more sharply, and $JUMP$ falls. These comovements are consistent with demand-side HFT contributing to the pre-2020 decline in information acquisition and supply-side HFT contributing to the subsequent improvement.

**INSERT FIGURE 6 HERE**

[11] Because year-end earnings are announced in the early months of the following year, 2022 is the last year with complete announcement coverage in our sample. The $JUMP$ analysis therefore covers announcements from 2010 through 2022.

Taken together, the annual estimates and level patterns identify HFT composition as a plausible part of the time-series evidence. They do not rule out other explanations for the evolution of $JUMP$ and should not be read as a decomposition of the aggregate trend.

*6.2.HFT and liquidity*

We next examine trading costs as a mechanism linking HFT to information acquisition. If liquidity-demanding HFT raises trading costs while liquidity-supplying HFT lowers them, spreads should account for part of the opposing relationships with $JUMP$ and should themselves be related to the two HFT components in opposite directions.

We begin with a mediation-style test. Columns (v) and (vi) of Table 7 repeat the $JUMP$ specifications after removing effective spread from both stages. This comparison does not identify a formal causal mediation effect, but it shows whether conditioning on spreads materially changes the IV coefficients.

Without the spread control, the $HFT_D$ coefficient rises from 0.469 to 0.549, or by 17%, and the absolute $HFT_S$ coefficient rises from 0.229 to 0.306, or by 34%. A cluster bootstrap that resamples tickers 1,000 times rejects equality between the with-spread and without-spread estimates at the 1% level for both components. The larger change for $HFT_S$ is consistent with passive HFT being linked to information acquisition partly through quoted trading costs.

This evidence is suggestive rather than definitive: removing spreads can also change the specification through channels other than mediation. We therefore turn to the direct IV relationship between the HFT measures and quoted spreads and ask whether its sign and stability match the trading-cost interpretation.

The spread analysis extends prior evidence in two ways. It separates liquidity demand from supply, which aggregate HFT measures combine, and it follows both relationships over 2010-2023. The long panel also allows a focused test during COVID-19, when market-wide stress persisted for weeks rather than seconds or minutes.

We use the same 2SLS structure as for $JUMP$, instrumenting each HFT component separately with lagged log price. The dependent variable is the monthly relative quoted spread, and the sample contains all U.S.-listed common-stock months from 2010 through 2023.

**INSERT TABLE 9 HERE**

Table 9 reports strong first stages with the same opposing signs as in Table 7. The F-statistics are 1169.86 for $HFT_D$ and 705.15 for $HFT_S$. Differences in coefficient magnitude reflect the broader monthly sample.

In the second stage, instrumented $HFT_D$ is positively associated with quoted spreads (0.974, $t$ = 16.21), whereas instrumented $HFT_S$ is negatively associated with spreads (-0.559, $t$ = -14.29). A one-standard-deviation increase in the corresponding fitted value is associated with spreads that are 38% higher for $HFT_D$ and 28% lower for $HFT_S$, relative to the sample mean. These magnitudes are comparable to estimates in related settings (Hendershott et al. 2011; Brogaard et al. 2015; Van Kervel and Menkveld 2019), but the identifying assumptions and the construction of our measures warrant caution.

One concern is mechanical overlap: spread variables are inputs to the ML measures, and quoted spread is the outcome here. The 2SLS design limits the role of purely contemporaneous spread innovations because fitted HFT variation comes from lagged price conditional on lagged controls and time fixed effects, but persistent spread-related information could still influence the fitted measures. We therefore retrain the models without spread inputs and repeat the spread regressions. We report these results in Online Appendix Table OA.F.1. The signs and qualitative conclusions are unchanged. This robustness check reduces, although does not eliminate, the concern; hence we do not interpret the magnitudes as mechanical-free causal effects.

The opposing signs match the economic roles assigned to the measures. Speed can help liquidity suppliers manage inventory and update stale quotes, reducing inventory and adverse-selection costs (Menkveld 2013; Baldauf and Mollner 2020; Aït-Sahalia and Sağlam 2024; Jovanovic and Menkveld 2025). Aggressive fast traders can instead pick off stale quotes or trade quickly on news, leading other liquidity providers to widen quotes (Hoffmann 2014; Biais et al. 2015; Budish et al. 2015; Foucault et al. 2016). The estimates are thus consistent with a nuanced view: the liquidity associated with HFT depends on which of passive and aggressive strategies dominate the trading environment.

Next, we re-estimate the spread specifications by year to assess whether these relationships are confined to a particular part of the sample.

**INSERT FIGURE 7 HERE**

Figure 7 shows the same opposing signs in every year from 2010 to 2023, with 90% confidence intervals excluding zero throughout. Magnitudes are not constant: the positive $HFT_D$ coefficient rises from about 0.55 in 2010 to roughly 1.7 in 2020, while the negative $HFT_S$ coefficient grows in absolute value from about 0.25 to about 0.95 by 2020-2021 before moderating. The annual evidence therefore supports sign stability, not an invariant coefficient. COVID-19 provides a sharper test of liquidity supply under prolonged market-wide stress. Earlier evidence emphasizes short-lived episodes such as the Flash Crash, in which fast liquidity may not absorb extreme order imbalances (Kirilenko et al. 2017; Brogaard et al. 2018). The pandemic differs because elevated volatility persisted long enough for traders to adjust inventories and capital allocation.

**INSERT FIGURE 8 HERE**

Figure 8 plots weekly averages from November 2019 through June 2020. Quoted spreads roughly triple in March and then normalize by May. $HFT_D$ rises about 5% from its pre-COVID average to its peak, while $HFT_S$ rises about 13%. The descriptive evidence therefore shows no withdrawal of measured passive HFT at the crisis peak; spreads nevertheless widen sharply, indicating that any additional HFT supply was small relative to the market-wide liquidity shock.

Table 10 evaluates whether the spread associations change during the crisis. We estimate daily 2SLS regressions over three months before, February-March 2020, and three months after the crisis window. Lagged daily log price instruments each HFT component, and the fitted HFT measure is interacted with the COVID indicator. Day fixed effects absorb the standalone indicator.

**INSERT TABLE 10 HERE**

The first stages retain the usual opposing signs. In the baseline daily second stages, $HFT_D$ is associated with wider spreads and $HFT_S$ with narrower spreads, consistent with the monthly estimates.

The interaction terms' estimates are asymmetric. The $HFT_D$ interaction term is -0.005 ($t$ = -0.14), providing no evidence that its spread association changed during COVID. The $HFT_S$ interaction term is -0.138 ($t$ = -3.70); hence, the negative coefficient moves from -0.488 outside the crisis window to -0.625 during it, a 28% increase in absolute magnitude. Conditional on the IV assumptions, this result indicates a stronger negative association between passive HFT and spreads during the crisis.

The result is consistent with fast liquidity suppliers responding to higher expected compensation when volatility rises. Some suppliers may withdraw, increasing liquidity rents for those that remain. Fast traders can adjust quotes and inventories more rapidly and, over a sustained episode, can redeploy capital toward market making. Models in which the relative payoff to fast liquidity provision rises with volatility generate this pattern (Nagel 2012; Hoffmann 2014; Bernales 2019).

Chakrabarty and Pascual (2023) likewise find that algorithmic traders did not disproportionately withdraw liquidity supply during COVID-19, although their aggregate measures cannot separate passive from aggressive activity. Our decomposition locates the resilience on the passive side: $HFT_S$ increases and its negative spread association strengthens. Because lower trading costs can support the return to informed trading, this pattern is also consistent with the improvement in pre-announcement information acquisition in 2020, while falling short of establishing that HFT caused that improvement.

*6.3.HFT and informed trading*

The results thus far link HFT composition, trading costs, and information acquisition. If the two HFT components are related to the cost of trading on information in opposite directions, they would also have opposing relationships with informed trading itself. We examine this final link using the informed trading intensity (ITI) measures of Bogousslavsky et al. (2024), expecting $HFT_D$ to be negatively associated with informed trading and $HFT_S$ to be positively associated.

Bogousslavsky et al. (2024) construct stock-day measures of informed trading intensity from trades disclosed in Schedule 13D filings. $ITI(13D)$ uses the full 60-day filing window; $ITI(patient)$ uses the first 40 days, when activists build positions gradually; and $ITI(impatient)$ uses the final 20 days, when trading becomes more aggressive. The distinction is useful because patient informed traders interact repeatedly with the limit order book and may therefore be more exposed to both adverse selection from aggressive HFT and liquidity supplied by passive HFT (Kaniel and Liu 2006).

**INSERT TABLE 11 HERE**

Table 11 reports second-stage estimates from the same 2SLS framework used for $JUMP$ and spreads. For $ITI(13D)$, instrumented $HFT_D$ is negatively associated with informed trading (-0.476, $t$ = -9.81), whereas instrumented $HFT_S$ is positively associated with it (0.184,

t = 8.65). A one-standard-deviation increase in the corresponding fitted value is associated with 14% lower $ITI(13D)$ for $HFT_D$, and 6% higher $ITI(13D)$ for $HFT_S$.

The opposing relationships are concentrated among patient informed traders. The coefficients for $ITI(patient)$ are -0.701 ($t$ = -12.30) for $HFT_D$ and 0.271 ($t$ = 9.85) for $HFT_S$, corresponding to 21% lower and 9% higher $ITI(patient)$, respectively. For $ITI(impatient)$, the coefficients are -0.077 and 0.030, with economic magnitudes of only 2% and 1% of the mean.

This concentration is consistent with the trading-cost channel. Patient informed traders split orders over time and use limit orders, repeatedly exposing them to being picked off by aggressive HFT while also allowing them to benefit from depth and tighter spreads supplied by fast market makers. Impatient traders cross the spread and complete their trades more quickly, limiting their exposure to both channels. As elsewhere, the IV interpretation depends on the identifying assumptions and does not establish a structural causal effect.

Across the three applications, liquidity-supplying and liquidity-demanding HFT have opposing relationships with market quality. $HFT_S$ is associated with narrower spreads, more pre-announcement information acquisition, and more informed trading; $HFT_D$ is associated with the reverse. The signs are stable over time. During COVID-19, the negative association between $HFT_S$ and spreads strengthens while the $HFT_D$ association is unchanged, evidence consistent with resilient passive HFT under prolonged stress.

## 7. Extensions and further robustness analyses

This section summarizes six robustness questions. Each test addresses a specific concern with the implementation details reported in the cited Online Appendix sections, where applicable.

First, do the measures respond around public information releases? Figure OA.C.1 shows that both HFT components rise around scheduled earnings announcements and peak on the announcement day. Relative to pre-announcement levels, $HFT_S$ rises 6.3% and $HFT_D$ rises 2.8% over the three-day event window.

Figure OA.C.2 reports the corresponding test for unscheduled M&A announcements. In contrast to scheduled earnings news, both measures begin rising only at or immediately before the announcement. HFT activity builds ahead of events on a known calendar but not ahead of unpredictable ones, consistent with HFT responding to public information rather than trading on private knowledge of upcoming news (Budish et al. 2015; Aquilina et al. 2022).

Second, can the information-acquisition results be obtained using the small (labeled) HFT sample, and are our findings sensitive to temporal distance from the 2009 training period? Repeating the $JUMP$ regressions with the 120 Nasdaq stocks produces no statistically significant estimates, likely due to the fact that there are too few earnings events in the sample for this low-frequency outcome (Table OA.D.1), underscoring the value of the broader coverage provided by our measures. Restricting the long-panel analysis to 2010-2012 addresses the temporal-distance concern and yields the same opposing signs: $HFT_D$ is associated with less pre-announcement information acquisition and $HFT_S$ with more (Table OA.D.2).

Third, do the separate $HFT_D$ and $HFT_S$ equations mask their joint relationship with the outcomes? Online Appendix E uses Lewbel's (2012) heteroskedasticity-based instruments to include both endogenous components in the same equation. Table OA.E.1 shows that the joint estimates retain the baseline signs for $JUMP$, quoted spread, and $ITI(13D)$. Wald tests indicate a stronger demand-side relationship for all three outcomes.

Fourth, do richer quote-level inputs materially improve measurement? We add message counts, quote updates, sub-100-share volume, and 100-millisecond midpoint variation to the baseline inputs to train the ML models on January-June 2009 data, and use the resulting HFT measures to predict the July-December 2009 data. The additional inputs raise the baseline out-of-sample $R^2$ only from 82% to 84%. The augmented and baseline measures have a correlation coefficient of 0.99 for $HFT_S$ and 0.96 for $HFT_D$, and their validation regressions are similar. Thus, the accessible daily inputs capture most of the predictive content in the more granular quote data.

Fifth, could spread-related inputs in the ML feature set mechanically generate the estimated relationships between our HFT measures and quoted spreads? We retrain our ML model after excluding quoted spread, effective spread, realized spread, and price impact from the model's feature set, reconstruct the liquidity-demanding and liquidity-supplying HFT measures, and re-estimate the 2SLS regressions relating quoted spreads to the instrumented HFT measures. The estimates have the same signs and similar magnitudes (Table OA.F.1).

Sixth, could scaling HFT volume by total volume drive the results? We retrain the model using unscaled liquidity-demanding and liquidity-supplying HFT volume as the outcomes. The unscaled measures continue to outperform conventional proxies and respond in the predicted directions to the technology and latency-arbitrage tests (Tables OA.G.1-OA.G.3).

## 8. Conclusion

Studying HFT is difficult because datasets that separately identify liquidity-demanding and liquidity-supplying activity are scarce. The few proprietary datasets that provide this distinction cover narrow cross-sections of stocks, short periods or specific venues. While public-data proxies provide broader coverage, they do not distinguish liquidity supply from liquidity demand. We bridge this gap by training tree-based ensemble models on proprietary Nasdaq data and public intraday variables to construct separate daily measures of liquidity-supplying and liquidity-demanding HFT for U.S. stocks from 2010 to 2023. The measures add information beyond standard proxies and, importantly, capture within-stock time-series variation that those proxies largely miss. We make the resulting stock-day measures publicly available for future research.

A central concern is whether the mapping between public intraday variables and HFT activity is stable across markets and persists over time. Our validation provides several pieces of evidence. The measures respond in the predicted directions to U.S. market-structure changes and latency-arbitrage opportunities. In proprietary Euronext Paris data, similar public signals identify HFT across markets, and a model trained on 2009 Euronext observations remains predictive of observed HFT activity in 2017. Together, these results reduce the concern that the mapping reflects only one market or a short training period.

Separating HFT liquidity supply from demand changes how we interpret its relationship with information production and market quality. Liquidity-supplying HFT is linked with greater pre-announcement information acquisition, more informed trading, and narrower spreads, while liquidity-demanding HFT shows the opposite associations. These signs persist over the long panel. During COVID-19, liquidity-supplying HFT increased and its negative association with spreads strengthened, evidence consistent with resilient passive HFT during prolonged market stress. The informed-trading relationships are concentrated among patient traders who build positions gradually through the limit order book. Finally, the public proxies used in the information-acquisition literature are more closely related to liquidity-demanding HFT, helping reconcile our results with earlier evidence that aggregate algorithmic trading reduces information acquisition. More broadly, the results show why treating HFT as a single activity can obscure economically distinct channels.

**Declaration of generative AI and AI-assisted technologies in the writing process**

During the preparation of this work, the authors used multiple LLMs from Anthropic, Google, and OpenAI to improve writing clarity and style. After using this tool/service, the authors reviewed and edited the content as needed and take full responsibility for the content of the publication.

**Figure 1**
**Feature importance plot.**
This figure reports impurity-based feature importance for the regression-tree ensemble. For each fitted model, a feature's importance is its normalized, observation-weighted total reduction in squared-error impurity across splits that use the feature. Bars report the mean importance across trees. Importance measures predictive contribution within the fitted model; it is not a test of statistical significance or causality.

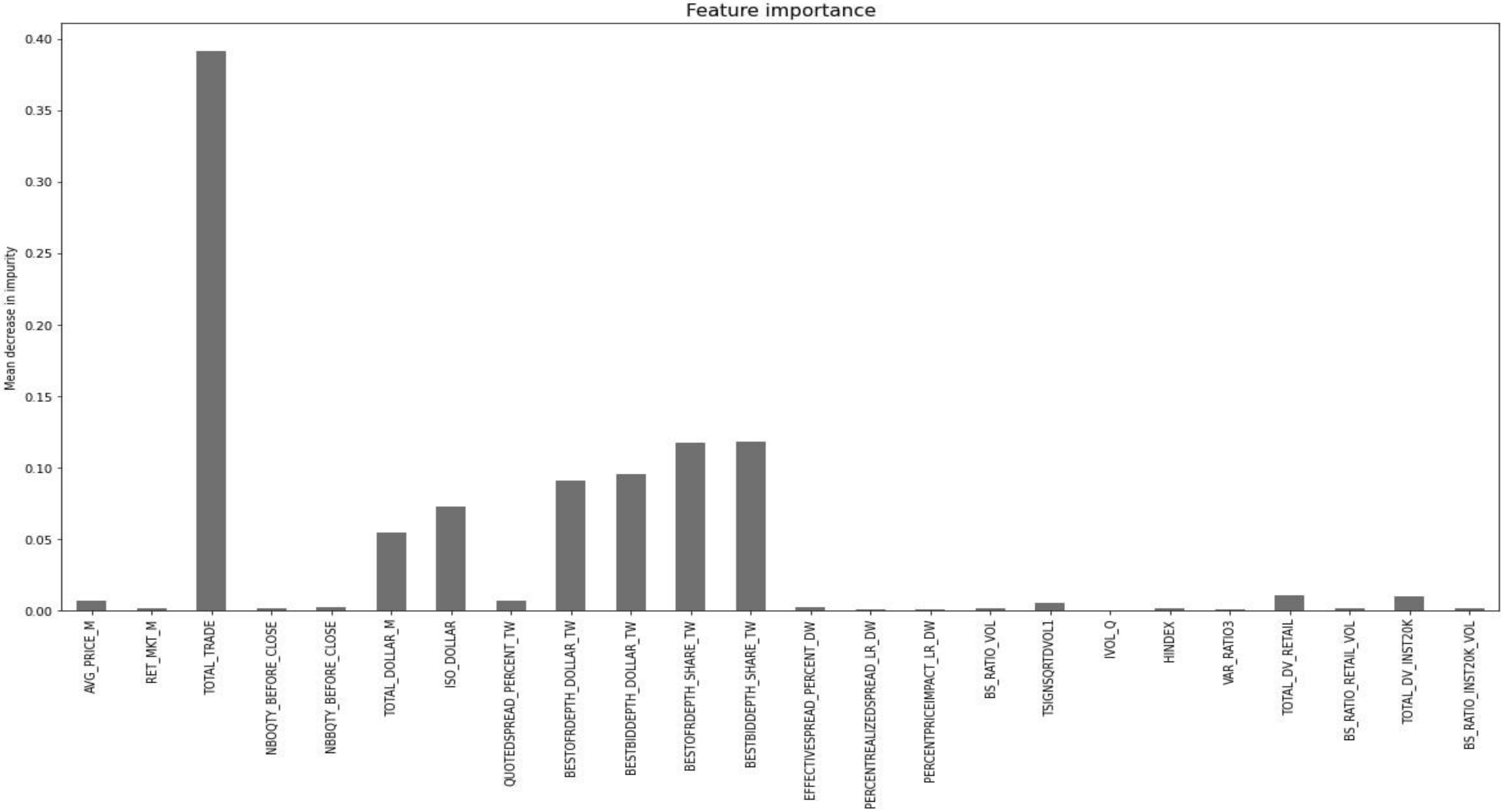

**Figure 2**

**Partial dependence plots of ML-generated HFT proxies on selected variables.**

This figure reports partial-dependence plots for selected inputs. At each displayed input value, the curve is the model's average predicted HFT share after averaging over the observed values of the other inputs. The plots summarize the fitted model's functional form and may reveal nonlinearity; they describe fitted associations and do not identify causal effects.

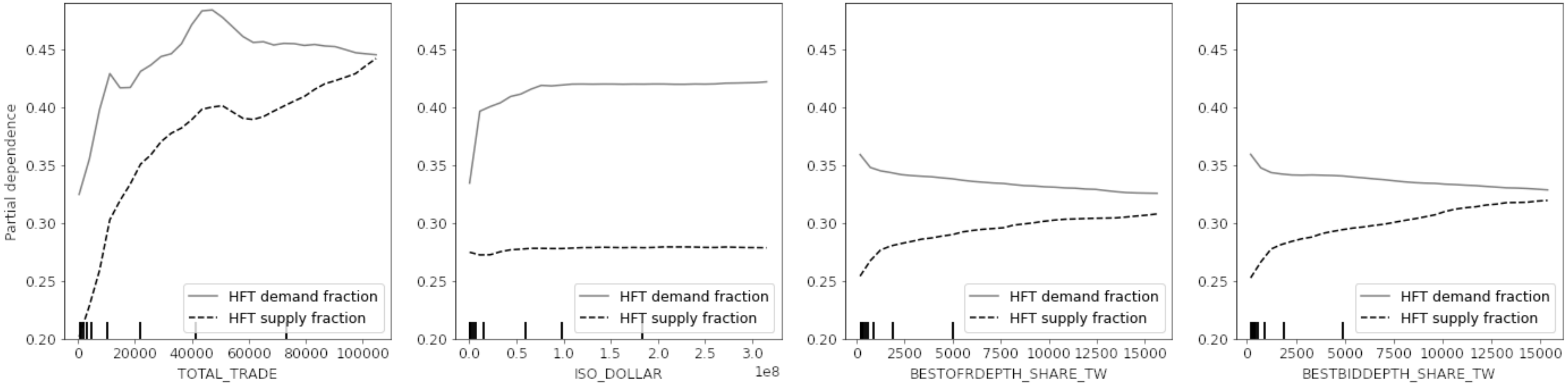

**Figure 3**
**Feature importance for 2009 Nasdaq and 2009 Euronext.**
This figure compares impurity-based feature importance from extremely randomized regression-tree models trained separately on 2009 Nasdaq data (green) and 2009 Euronext data (blue). A feature's importance is its normalized, observation-weighted total reduction in squared-error impurity across splits that use the feature. Panel A reports individual variables. Panel B groups variables by economic category and reports mean importance within each group. Importance measures predictive contribution within each fitted model, not statistical significance or causality.

Panel A: Individual variable level

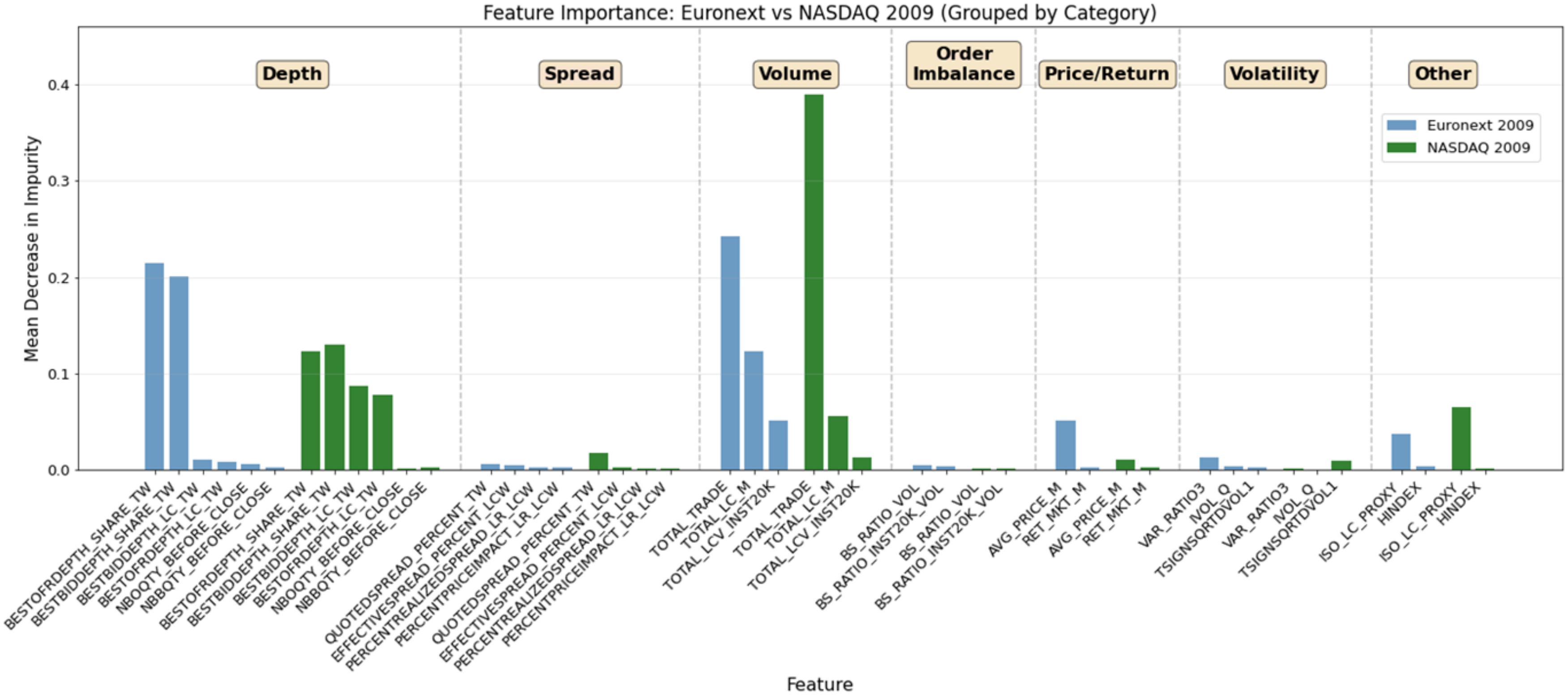

Panel B: Group level

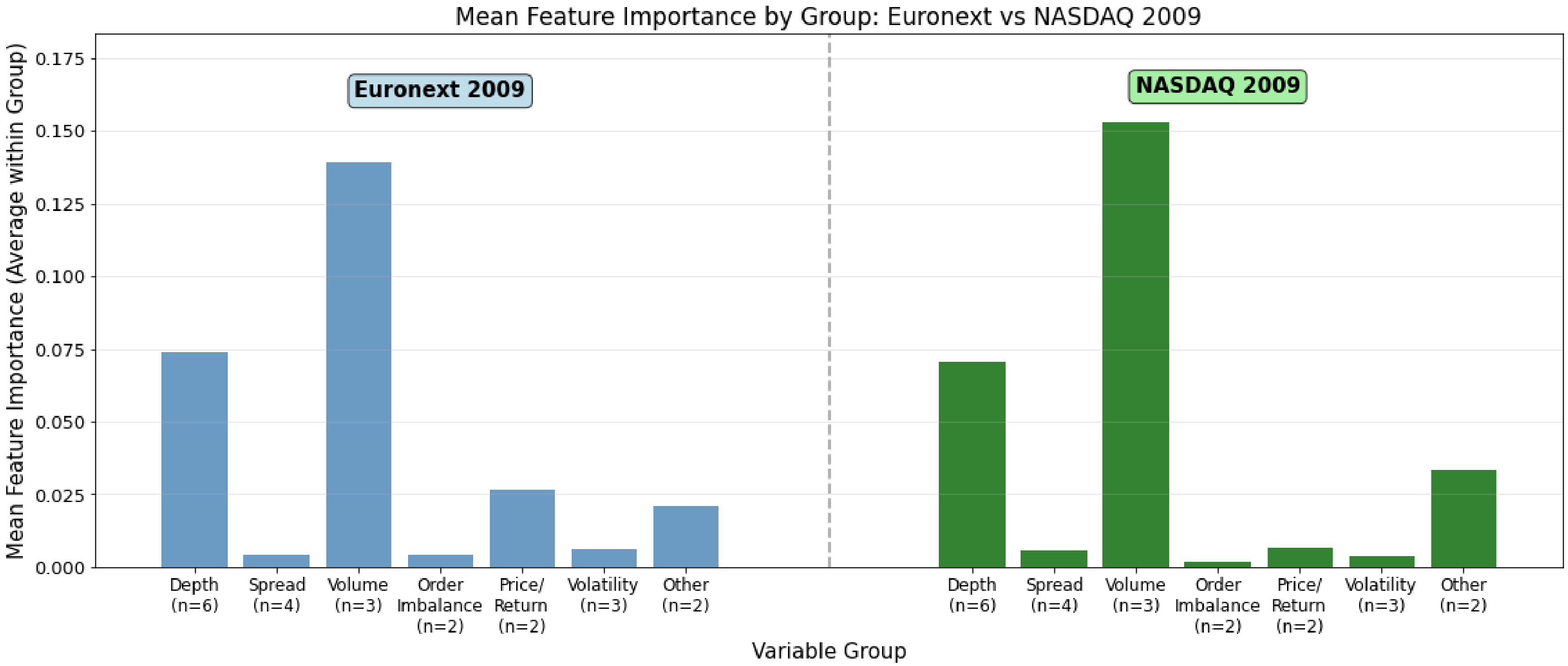

**Figure 4**
**Feature importance for 2009 and 2017 Euronext.**
This figure compares impurity-based feature importance from extremely randomized regression-tree models trained separately on 2009 Euronext data (blue) and 2017 Euronext data (green). A feature's importance is its normalized, observation-weighted total reduction in squared-error impurity across splits that use the feature. Panel A reports individual variables. Panel B groups variables by economic category and reports mean importance within each group. Importance measures predictive contribution within each fitted model, not statistical significance or causality.

Panel A: Individual variable level

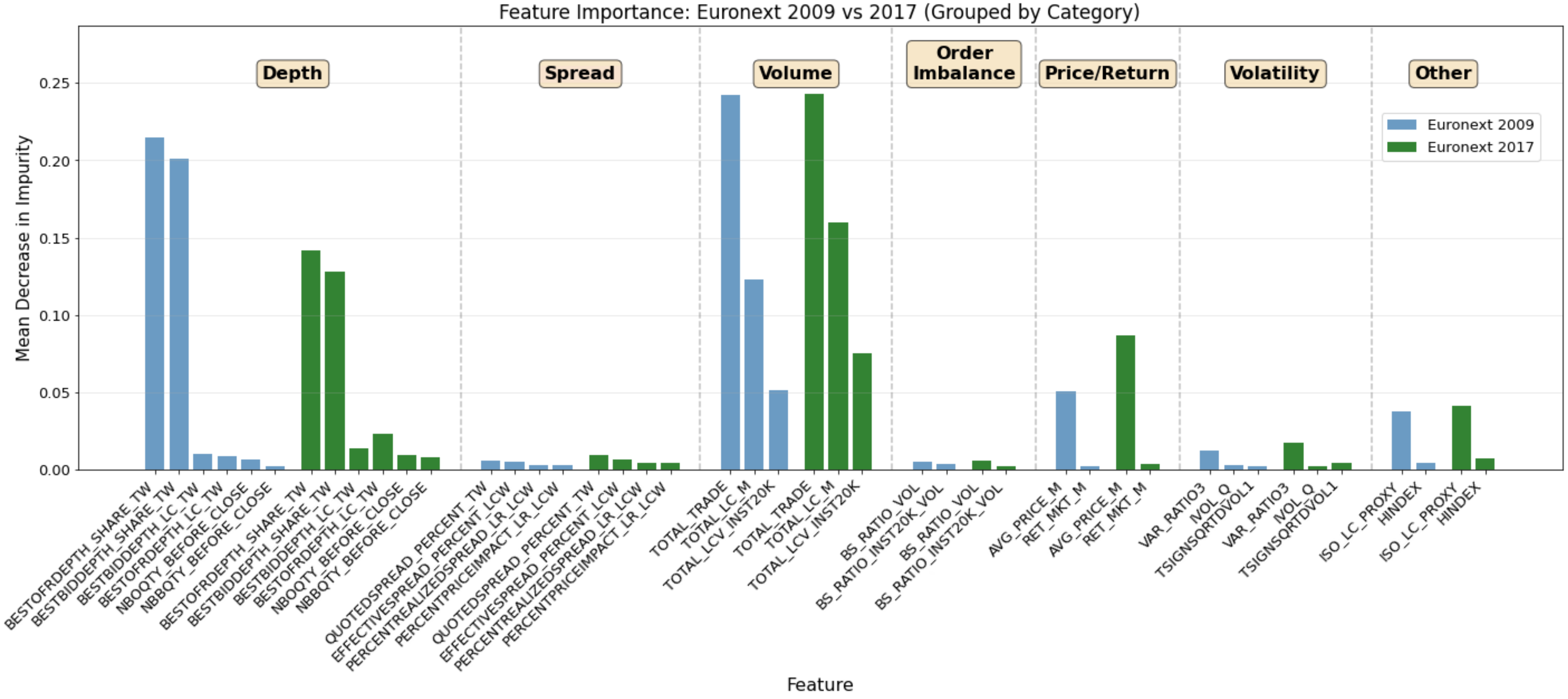

Panel B: Group level

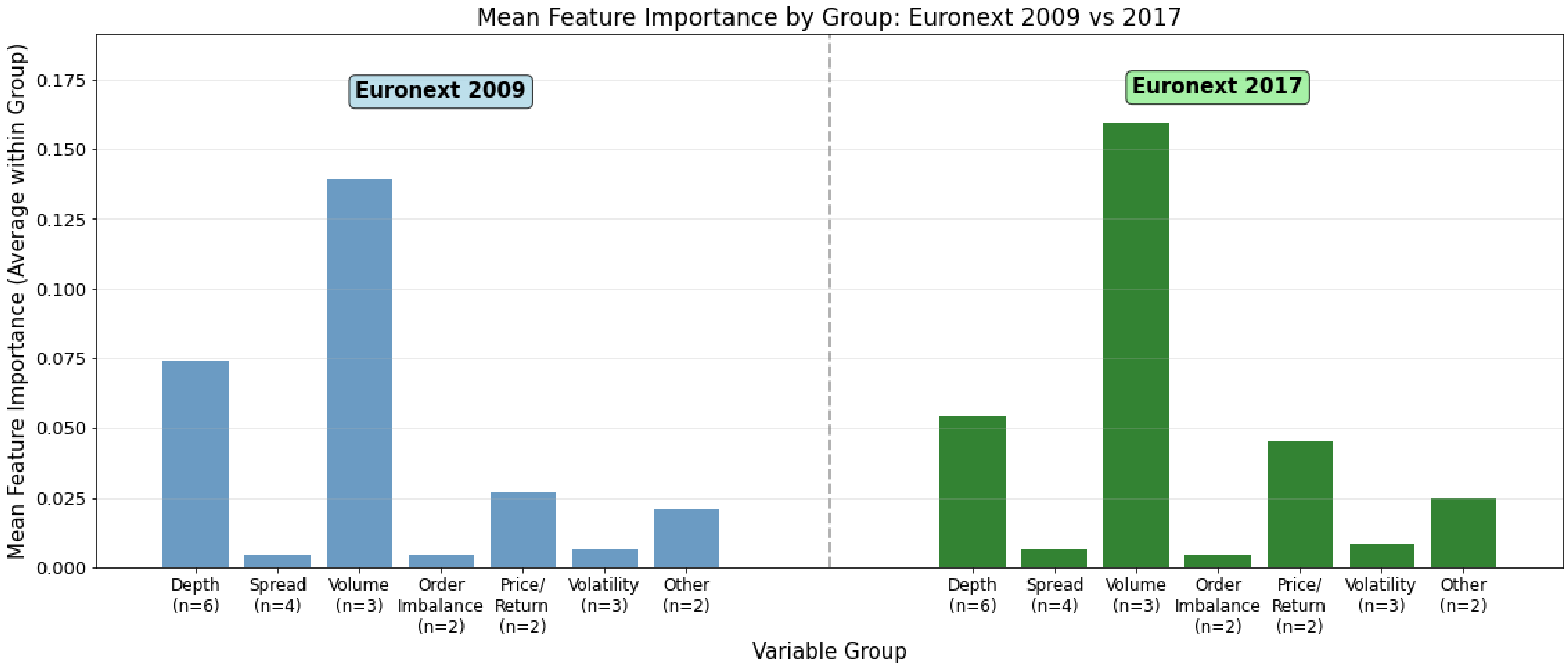

**Figure 5**
**Time-series of the relation between HFT and JUMP.**
Panel A plots the annual mean and median of $JUMP$, the price jump ratio. $JUMP$ measures information acquisition for stock $i$ as the ratio of cumulative abnormal returns over [-1, 1] to cumulative abnormal returns over [-21, 1] around quarterly earnings announcements. Panel B plots the annual IV coefficients on $HFT_D$ and $HFT_S$ from year-by-year estimation, with 90% confidence intervals. The sample includes US-listed common stocks from 2010 to 2022.

Panel A: The annual mean and median of $JUMP$

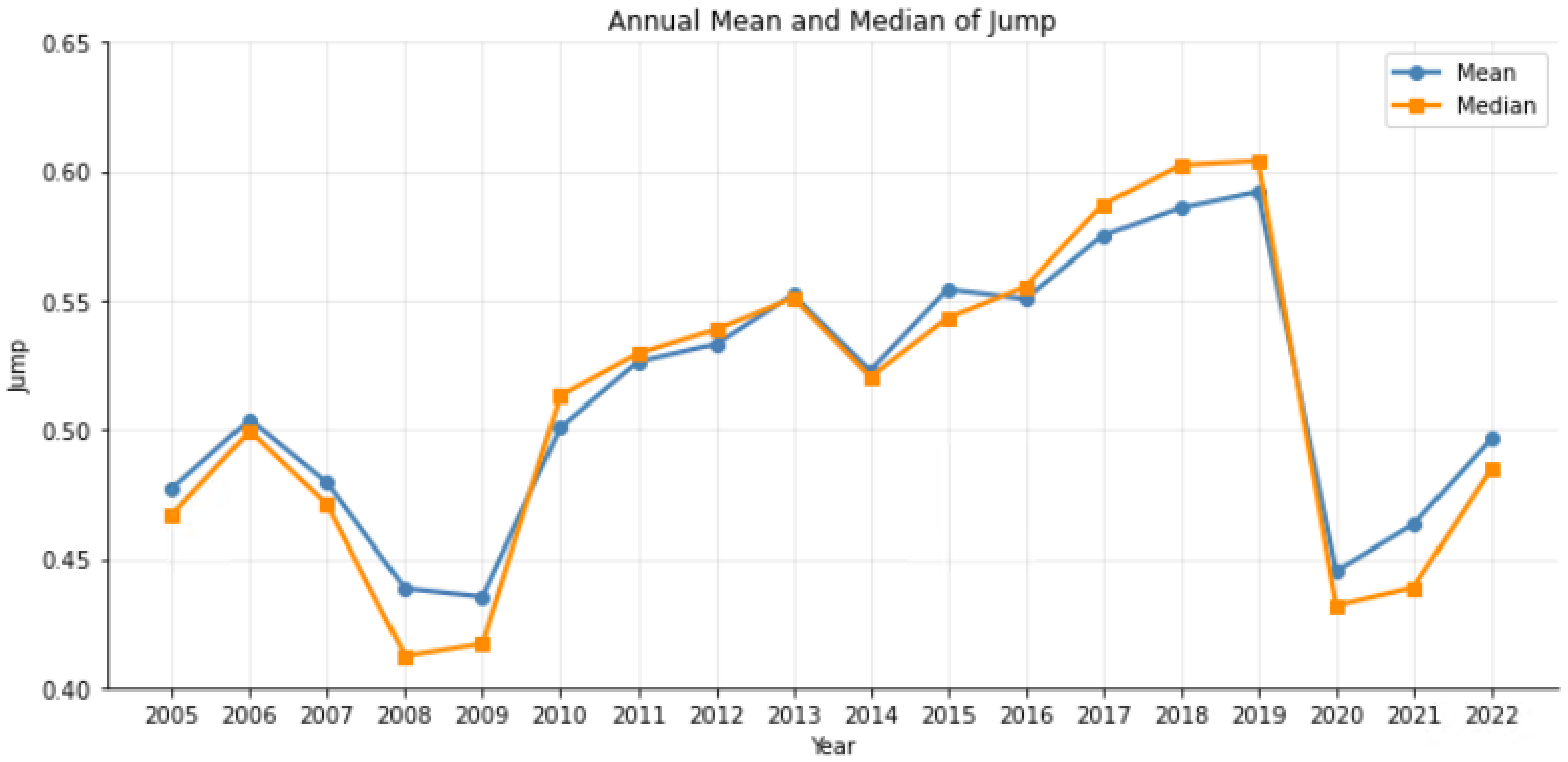

Panel B: The annual IV coefficients on $HFT_D$ and $HFT_S$, with 90% confidence intervals.

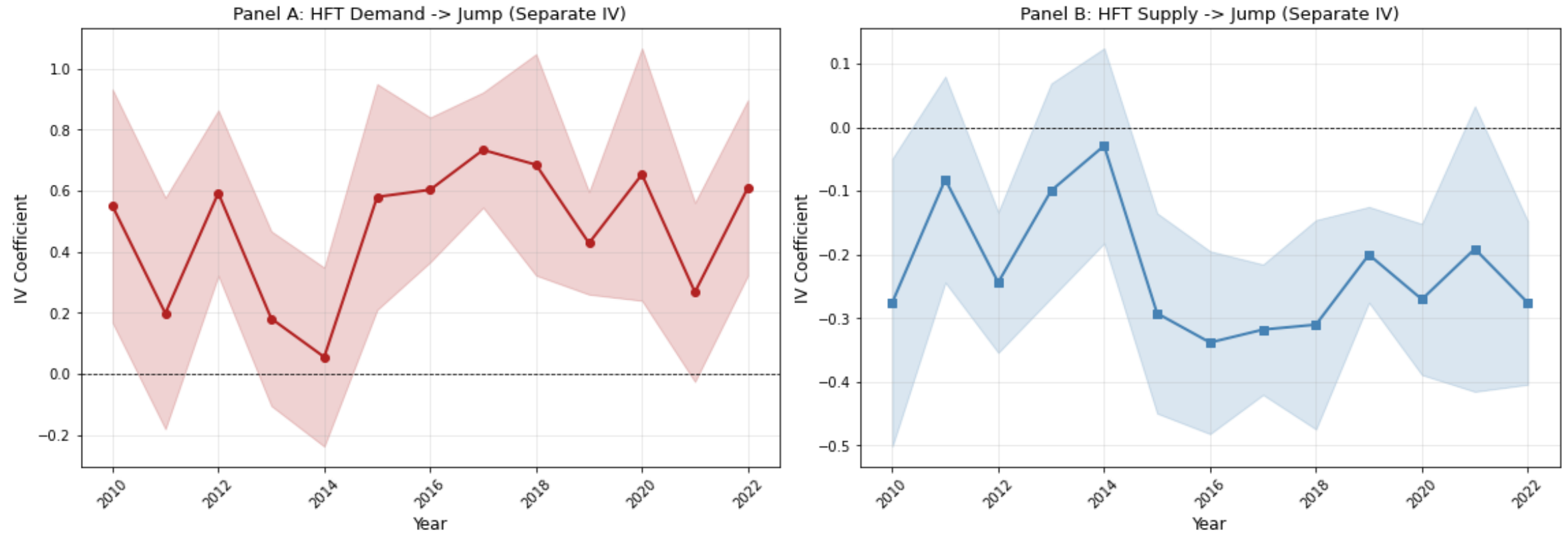

**Figure 6**
**Time-series of JUMP, HFT demand, and HFT supply.**
This figure plots annual cross-sectional means of $JUMP$ alongside the annual cross-sectional means of $HFT_D$ and $HFT_S$ from 2010 through 2022. $JUMP$ is the price jump ratio, measured for stock $i$ as the ratio of cumulative abnormal returns over [-1, 1] to cumulative abnormal returns over [-21, 1] around quarterly earnings announcements. $HFT_D$ and $HFT_S$ are our liquidity-demanding and liquidity-supplying HFT measures. Panel A plots $JUMP$ alongside $HFT_D$. Panel B plots $JUMP$ alongside $HFT_S$.

Panel A: The annual means of $JUMP$ and $HFT_D$

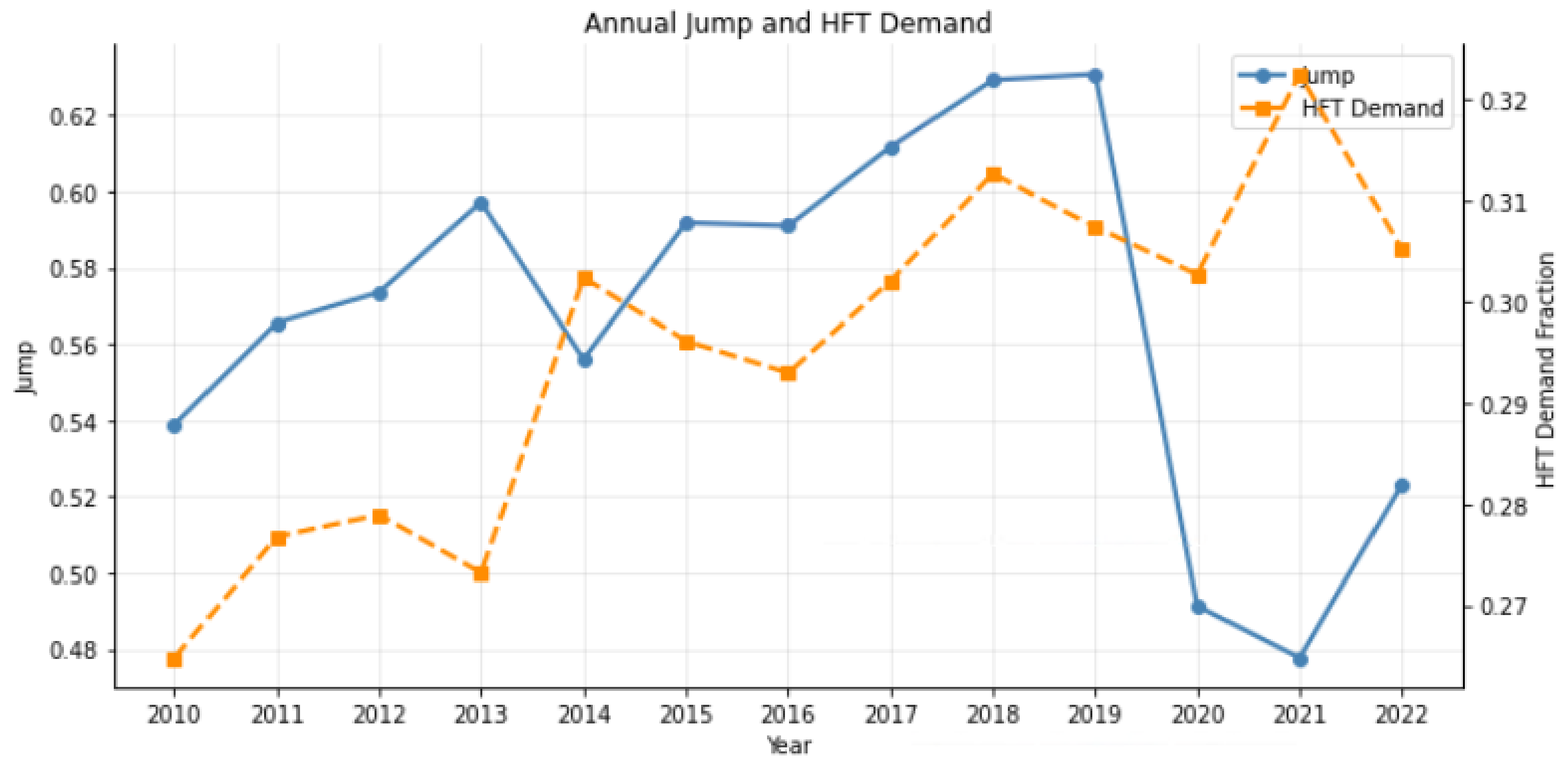

Panel B: The annual means of $JUMP$ and $HFT_S$

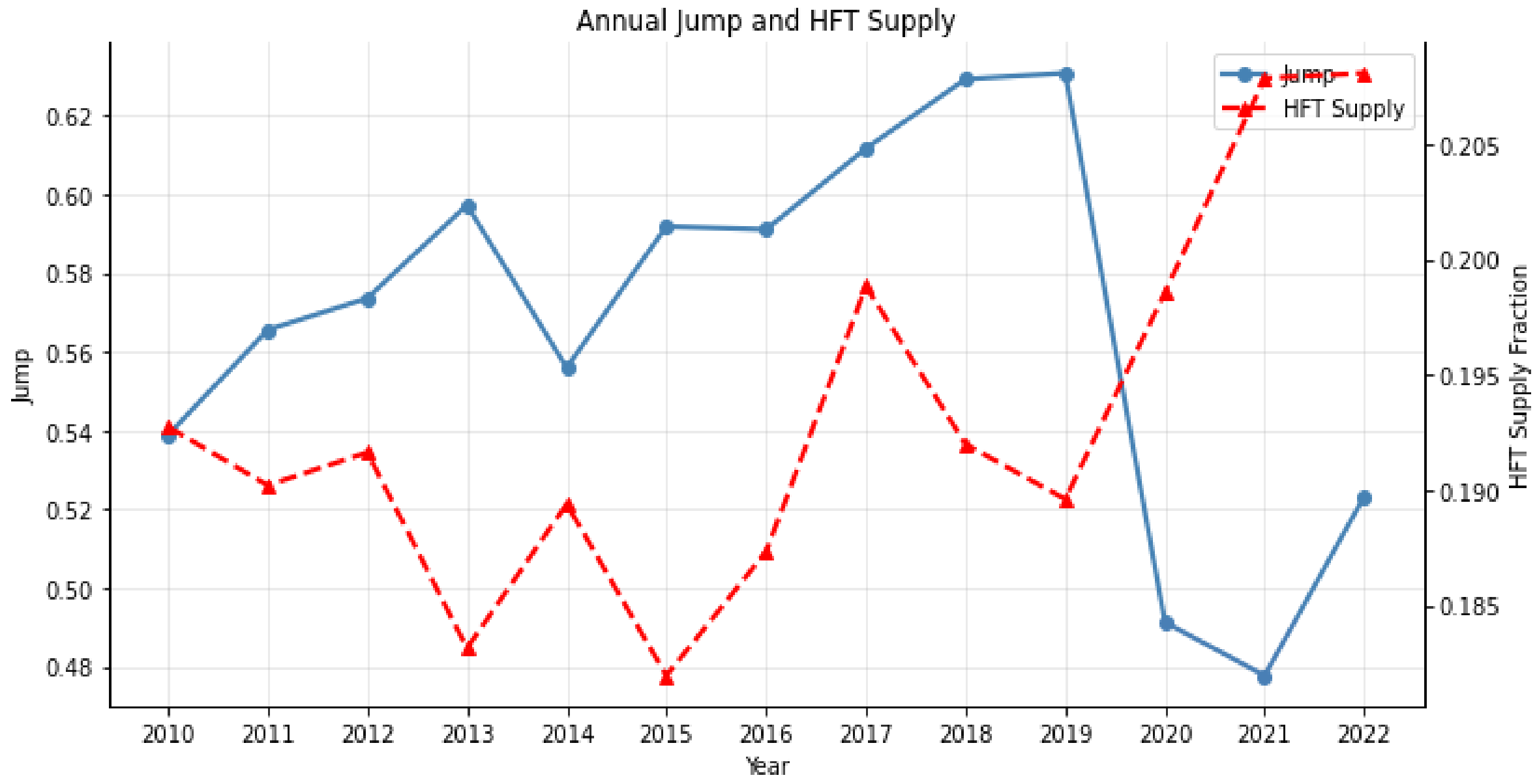

**Figure 7**

**Time-series of the relation between HFT and quoted spreads.**

This figure plots the annual IV coefficients on $HFT_D$ and $HFT_S$ from year-by-year estimation of the spread regression, with 90% confidence intervals from 2010 through 2023. $HFT_D$ and $HFT_S$ are our liquidity-demanding and liquidity-supplying HFT measures. Left panel plots the annual IV coefficient on $HFT_D$. Right panel plots the annual IV coefficient on $HFT_S$.

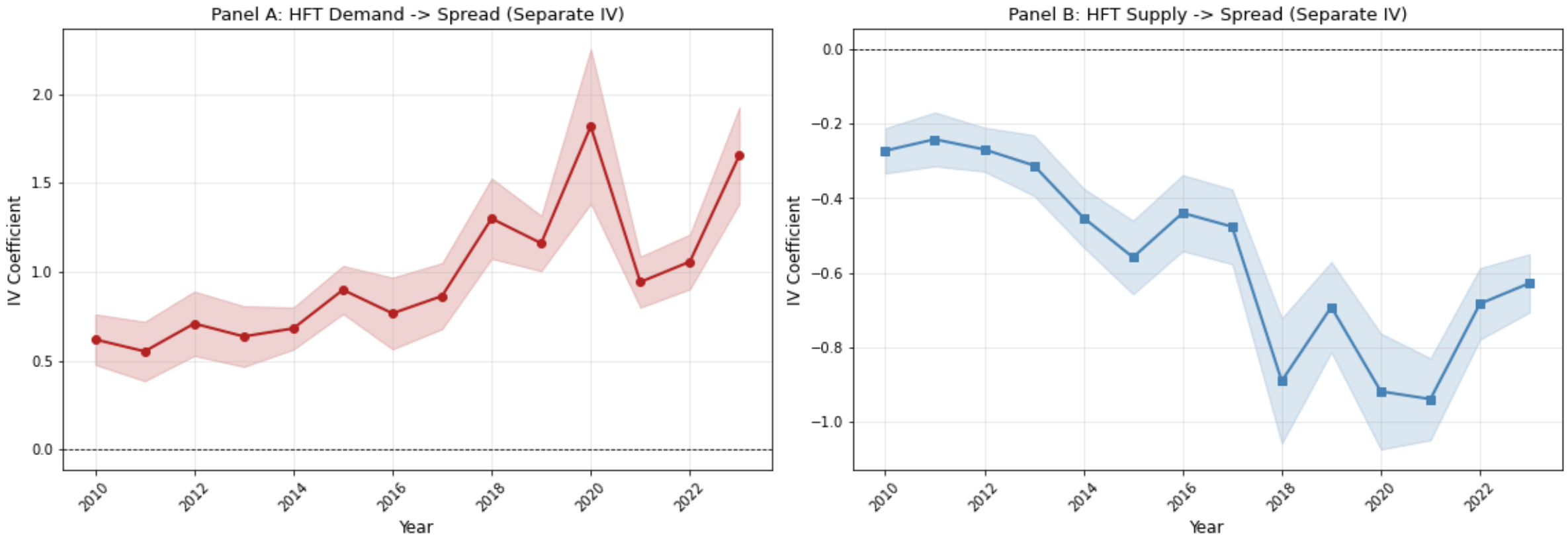

**Figure 8**

**Quoted spreads, HFT demand, and HFT supply around COVID-19.**

This figure plots weekly averages of the relative quoted spread, liquidity-demanding HFT, and liquidity-supplying HFT from November 2019 to June 2020. The shaded area marks the COVID-19 recession period as identified by the Federal Reserve, from February to April 2020.

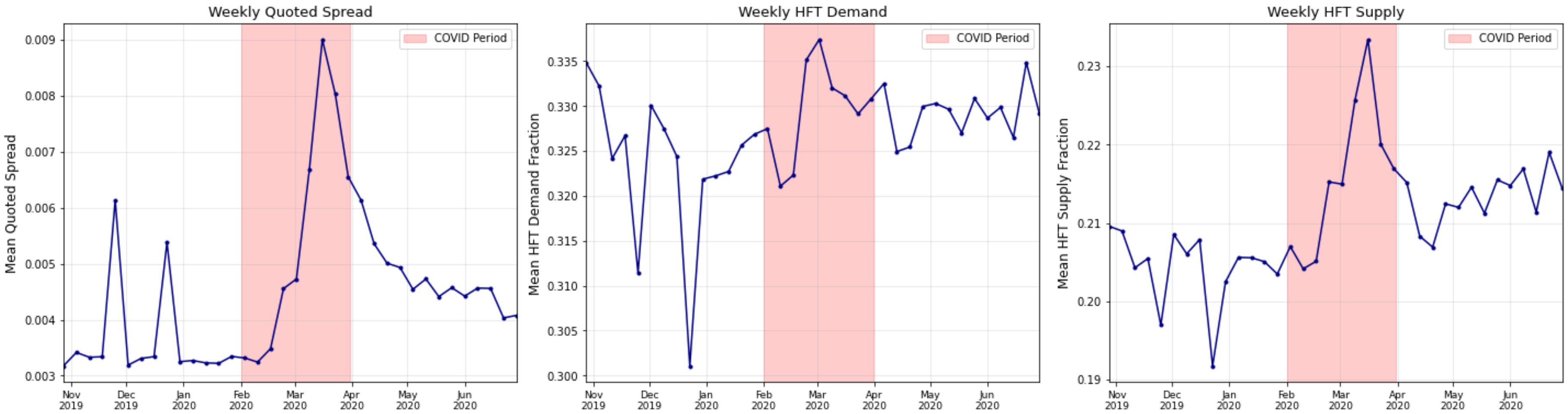

**Table 1**
**Input and output variables in the ML model training process.**
This table presents the variables used to train the ML model, including their notation, descriptions, and data sources. Panel A contains output variables from Nasdaq HFT data. Panel B details input variables derived from the TAQ database, with variable labels matching the WRDS TAQ Data Manual for easy reference.

| Variable | Description | Data source |
|---|---|---|
| *Panel A: Output variables used in the ML model.* | | |
| $NASD_HFT_D_{i,t}$ | Liquidity-demanding HFT share for stock $i$ on day $t$: daily shares in HH and HN trades divided by total shares in HH, HN, NH, and NN trades. | Nasdaq HFT |
| $NASD_HFT_S_{i,t}$ | Liquidity-supplying HFT share for stock $i$ on day $t$: daily shares in HH and NH trades divided by total shares in HH, HN, NH, and NN trades. | Nasdaq HFT |
| *Panel B: Input variables (features) used in the ML model.* | | |
| $AVG_PRICE_M_{i,t}$ | Average trade price during market hours (Open to Close) for stock $i$ in day $t$. | TAQ |
| $RET_MKT_M_{i,t}$ | Open to close return for stock $i$ in day $t$ is computed as the log return of the official opening price over the official closing price. | TAQ |
| $TOTAL_TRADE_{i,t}$ | The total number of trades for stock $i$ in day $t$. | TAQ |
| $NBOQTY_BEFORE_CLOSE_{i,t}$ | The best offer size of the last quote before market close for stock $i$ in day $t$. | TAQ |
| $NBBQTY_BEFORE_CLOSE_{i,t}$ | The best bid size of the last quote before market close for stock $i$ in day $t$. | TAQ |
| $TOTAL_DOLLAR_M_{i,t}$ | The total trade value in dollars during market hours for stock $i$ in day $t$. | TAQ |
| $ISO_DOLLAR_{i,t}$ | The sum of intermarket sweep order trade dollar value (during market hours) for stock $i$ in day $t$. | TAQ |
| $QUOTEDSPREAD_PERCENT_TW_{i,t}$ | The time-weighted percentage quoted spread (during market hours) for stock $i$ in day $t$. The quoted spread is calculated as the difference between ask and bid prices for each transaction divided by the mid-price (the average of ask and bid prices). | TAQ |
| $BESTOFRDEPTH_DOLLAR_TW_{i,t}$ | The time-weighted best offer dollar depth (during market hours) for stock $i$ in day $t$ is determined based on the size of the best ask price. | TAQ |
| $BESTBIDDEPTH_DOLLAR_TW_{i,t}$ | The time-weighted best bid dollar depth (during market hours) for stock $i$ in day $t$ is determined as the size of the best bid price. | TAQ |
| $BESTOFRDEPTH_SHARE_TW_{i,t}$ | The time-weighted best offer share depth (during market hours) for stock $i$ in day $t$ is determined based on the size of the best ask price. | TAQ |
| $BESTBIDDEPTH_SHARE_TW_{i,t}$ | The time-weighted best bid share depth (during market hours) for stock $i$ in day $t$ is determined based on the size of the best bid price. | TAQ |
| $EFFECTIVESPREAD_PERCENT_DW_{i,t}$ | The dollar value-weighted percentage effective spread for stock $i$ in day $t$. The effective spread is calculated using the following equation: $Effective\ Spread = 2D_k(P_k - M_k)/M_k$, where $k$ denotes transaction, $D_k$ denotes the sign of transaction (-1 for sale and +1 for buy), $P_k$ is the transaction price, and $M_k$ is the prevailing mid-price for each transaction. Lee and Ready (1991) algorithm is used for trade classification. | TAQ |
| $PERCENTREALIZEDSPREAD_LR_DW_{i,t}$ | The dollar value-weighted percentage realized spread for stock $i$ in day $t$. The realized spread is calculated using the following equation: $Realized\ Spread = 2D_k(P_k - M_{k+5})/M_k$, where $M_{k+5}$ is the bid-ask mid-point five minutes after the $k$th trade, and all other variables are as previously defined. Lee and Ready (1991) algorithm is used for trade classification. | TAQ |
| $PERCENTPRICEIMPACT_LR_DW_{i,t}$ | The dollar value-weighted percentage price impact for stock $i$ in day $t$. The price impact is calculated using the following equation: $Percent\ Price\ Impact = 2D_k(M_{k+5} - M_k)/M_k$, where all variables are as previously defined. Lee and Ready (1991) algorithm is used for trade classification. | TAQ |
| $BS_RATIO_VOL_{i,t}$ | The absolute percentage order imbalance for stock $i$ in day $t$ is calculated as the absolute value of buy volume minus sell volume divided by the total trade volume. Lee and Ready (1991) algorithm is used for trade classification. | TAQ |
| $TSIGNSQRTDVOL1_{i,t}$ | The lambda (price impact coefficient) with intercept for stock $i$ in day $t$ is calculated using the following equation: $Ln\frac{M_{i,s}}{M_{i,s-300}} = \alpha + \lambda * \text{SSqrtDvol} + \epsilon$, where SSqrtDvol $= Sgn(\sum_{s-300}^{s} BuyDollar - \sum_{s-300}^{s} SellDollar) \times \sqrt{\lvert \sum_{s-300}^{s} BuyDollar - \sum_{s-300}^{s} SellDollar \rvert}$, where $M_{i,s}$ is the mid-price for stock $i$ at second $s$. | TAQ |
| $IVOL_Q_{i,t}$ | The quote-based intraday volatility for stock $i$ in day $t$ is calculated using the following equation: $Intraday\ Volatility = \frac{\sum_{s=1}^{S}(Ret_{i,s} - \overline{Ret_{i,s}})^2}{S-1}$, where $Ret_{i,s} = Ln\frac{M_{i,s}}{M_{i,s-1}}$ and $M_{i,s}$ is the mid-price for stock $i$ at second $s$. | TAQ |
| $HINDEX_{i,t}$ | The Herfindahl index calculated across 30-minute time units for stock $i$ in day $t$ is calculated using the following equation: $HIndex = \frac{\sum_{s=1}^{1800}\sum_{k=1}^{N}(P_k \times SHR_k)^2}{(\sum_{s=1}^{1800}\sum_{k=1}^{N} P_k \times SHR_k)^2}$, where $SHR_k$ is the shares of trade for transaction $k$. | TAQ |
| $VAR_RATIO3_{i,t}$ | The variance ratio for stock $i$ in day $t$ is calculated using the following equation: $Variance\ Ratio = \left|\frac{Var(Ret_{300t})}{5\times Var(Ret_{60t})} - 1\right|$, where $Var(Ret_{300t})$ is the variance of 5-minute log returns. | TAQ |
| $TOTAL_DV_RETAIL_{i,t}$ | The total dollar value of retail trades for stock $i$ in day $t$. Retail trades are identified by using the methodology described in Boehmer et al. (2021b). | TAQ |
| $BS_RATIO_RETAIL_VOL_{i,t}$ | The absolute percentage order imbalance for retail trading volume for stock $i$ in day $t$. Retail trades are identified by using the methodology described in Boehmer et al. (2021b). | TAQ |
| $TOTAL_DV_INST20K_{i,t}$ | The total dollar value of $20,000 institutional trades for stock $i$ in day $t$. $20,000 cutoff is based on Lee and Radhakrishna (2000). | TAQ |
| $BS_RATIO_INST20K_VOL_{i,t}$ | The absolute percentage order imbalance for $20,000 institutional trades' trading volume for stock $i$ in day $t$. $20,000 cutoff is based on Lee and Radhakrishna (2000). | TAQ |

**Table 2**
**Regression variables and summary statistics.**
This table provides summary statistics and definitions of variables used in our regression analyses. Variable names in the first column are followed by their measurement units in parentheses. For variables used in multiple regressions with different frequencies (daily, quarterly, etc.), we report summary statistics corresponding to their first appearance in our analyses. All variables are winsorized at the 1st and 99th percentiles. For the original Nasdaq dataset variables ($NASD_HFT_D_{i,t}$ and $NASD_HFT_S_{i,t}$), the sample covers the year 2009 and includes 120 randomly selected Nasdaq- and NYSE-listed firms with Nasdaq HFT data. For all other variables, the sample includes all U.S.-listed common stocks from 2010 to 2023, except for $JUMP$, which is available through 2022.

| Variable | Definition | Mean | Std | Min | p.25 | p.50 | p.75 | Max |
|---|---|---|---|---|---|---|---|---|
| $NASD_HFT_D_{i,t}$ | Liquidity-demanding HFT share for stock $i$ on day $t$: daily shares in HH and HN trades divided by total shares in HH, HN, NH, and NN trades. | 0.331 | 0.160 | 0.013 | 0.202 | 0.342 | 0.453 | 0.662 |
| $NASD_HFT_S_{i,t}$ | Liquidity-supplying HFT share for stock $i$ on day $t$: daily shares in HH and NH trades divided by total shares in HH, HN, NH, and NN trades. | 0.250 | 0.169 | 0.010 | 0.110 | 0.206 | 0.375 | 0.636 |
| $HFT_D_{i,t}$ | The liquidity-demanding HFT activity for stock *i* on day *t*, estimated using the ML model outlined in Section 3. | 0.316 | 0.112 | 0.025 | 0.222 | 0.335 | 0.414 | 0.602 |
| $HFT_S_{i,t}$ | The liquidity-supplying HFT activity for stock *i* on day *t*, estimated using the ML model outlined in Section 3. | 0.208 | 0.101 | 0.036 | 0.131 | 0.174 | 0.259 | 0.626 |
| $Volatility_{i,t}$ (1/00,000) | Daily volatility for stock *i* on day *t*, measured as the standard deviation of transaction-level returns. | 0.008 | 0.018 | 0.000 | 0.000 | 0.001 | 0.007 | 0.123 |
| $Spread_{i,t}$ (%) | Daily average of transaction-level spreads for stock *i* on day *t*, where each transaction-level spread is calculated as (ask price - bid price)/(0.5 × (ask price + bid price)). | 0.142 | 0.154 | 0.012 | 0.037 | 0.090 | 0.189 | 0.885 |
| $InvPrice_{i,t}$ | The inverse of stock price for stock *i* on day *t*. | 0.039 | 0.050 | 0.001 | 0.013 | 0.024 | 0.047 | 0.344 |
| $Volume_{i,t}$ ($'000,000,00) | Daily trading volume in dollars for stock *i* on day *t*. | 2.614 | 6.305 | 0.007 | 0.070 | 0.330 | 2.556 | 47.392 |
| $NLAO_{i,t}$ (000) | The number of latency arbitrage opportunities for stock *i* on day *t*, identified using the methodology detailed in Section 4.2. | 0.068 | 0.169 | 0.001 | 0.006 | 0.017 | 0.047 | 1.211 |
| $Flick_{i,t}$ (0) | Quote volatility for stock *i* on day *t*, measured as the daily average of standard deviations of quote midpoints calculated over 100 ms intervals. | 6.942 | 42.24 | 0.000 | 0.009 | 0.021 | 0.086 | 365.523 |
| $OLV_{i,t}$ | Daily average of trades smaller than 100 shares for stock *i* on day *t*. | 3.040 | 12.47 | 0.000 | 0.000 | 0.000 | 1.000 | 80.000 |
| $QuoteInt_{i,t}$ (000,000) | Daily count of changes in best quotes or quote depth for stock *i* on day *t*. | 0.191 | 0.264 | 0.002 | 0.031 | 0.059 | 0.253 | 2.775 |
| $QT_{i,t}$ | The ratio of quoted shares to traded shares for stock *i* on day *t*. | 15.82 | 16.23 | 2.19 | 5.88 | 9.51 | 18.71 | 85.70 |
| $MG_{i,t}$ (000,000) | The total number of messages (trade and quote) for stock *i* on day *t*. | 2.111 | 2.864 | 0.078 | 0.332 | 0.643 | 2.853 | 12.637 |
| $JUMP_{i,m}$ | Information acquisition proxy for stock *i* in month *m*, measured as the ratio of cumulative abnormal returns over [-1, 1] to cumulative abnormal returns over [-21, 1] around earnings announcements. | 0.517 | 0.427 | -0.543 | 0.227 | 0.510 | 0.794 | 1.663 |
| $MValue_{i,m}$ ($'000,000,000) | Market value for stock $i$ in month $m$, calculated as the average of daily market values over [-42, -22] around earnings announcements, where daily market value is closing price times shares outstanding. | 0.567 | 1.652 | 0.001 | 0.024 | 0.089 | 0.330 | 12.474 |
| $OIB20k_{i,m}$ | Institutional order imbalance for stock $i$ in month $m$, measured as the order imbalance of trades exceeding $20,000 over [-42, -22] around earnings announcements, obtained from TAQ. | 0.351 | 0.183 | 0.050 | 0.200 | 0.333 | 0.494 | 0.763 |
| $CT_{i,m}$ | The natural logarithm of the cancel-to-trade ratio for stock $i$ in month $m$, where the ratio is calculated as the average of daily (cancel messages/trade messages) over [-21, -1] around earnings announcements, obtained from MIDAS database. | 0.507 | 0.540 | -0.548 | 0.150 | 0.462 | 0.810 | 2.227 |
| $OLR_{i,m}$ | The natural logarithm of the odd-lot ratio for stock $i$ in month $m$, where the ratio is calculated as the average of daily proportions of trades below 100 shares over [-21, -1] around earnings announcements, obtained from MIDAS database. | 1.202 | 0.664 | -0.430 | 0.777 | 1.288 | 1.735 | 2.212 |
| $TO_{i,m}$ | The natural logarithm of the trade-to-order ratio for stock $i$ in month $m$, where the ratio is calculated as the average of daily (executed shares/submitted shares) over [-21, -1] around earnings announcements, obtained from MIDAS database. | -1.064 | 0.639 | -2.972 | -1.450 | -1.017 | -0.628 | 0.194 |

**Table 3**
**Comparative analysis of HFT measures.**
This table evaluates our ML-generated HFT measures against alternative proxies using the following models:

$$NASD_HFT_S_{i,t} = \alpha_i + \beta_t + \gamma_1 HFT_S_{i,t} + \gamma_2 Flick_{i,t} + \gamma_3 OLV_{i,t} + \gamma_4 QuoteInt_{i,t} + \gamma_5 QT_{i,t} + \gamma_6 MG_{i,t} + \varepsilon_{i,t}$$

$$NASD_HFT_D_{i,t} = \alpha_i + \beta_t + \gamma_1 HFT_D_{i,t} + \gamma_2 Flick_{i,t} + \gamma_3 OLV_{i,t} + \gamma_4 QuoteInt_{i,t} + \gamma_5 QT_{i,t} + \gamma_6 MG_{i,t} + \varepsilon_{i,t}$$

$NASD_HFT_D$ and $NASD_HFT_S$ are the observed Nasdaq liquidity-demanding and liquidity-supplying HFT shares. $HFT_D$ and $HFT_S$ are the corresponding ML-generated predictions, trained on January-June 2009. The alternative TAQ proxies are quote volatility ($Flick$), sub-100-share volume ($OLV$), quote intensity ($QuoteInt$), the quote-to-trade ratio ($QT$), and message count ($MG$). The regressors are standardized, but the dependent variables are not. Panels A and B include stock and day fixed effects; Panels C and D include day fixed effects only. The sample covers July-December 2009 for 120 Nasdaq- and NYSE-listed stocks. Standard errors are double-clustered by stock and day, with $t$-statistics in brackets. *, **, and *** indicate significance at the 10%, 5%, and 1% levels. Reported $R^2$ values are within-$R^2$.

Panel A: $NASD_HFT_S_{i,t}$

| | (i) | (ii) | (iii) | (iv) | (v) | (vi) | (vii) |
|---|---|---|---|---|---|---|---|
| $HFT_S_{i,t}$ | 0.104*** | | | | | | 0.096*** |
| | (8.52) | | | | | | (7.52) |
| $Flick_{i,t}$ | | 0.002* | | | | | 0.001 |
| | | (1.79) | | | | | (1.00) |
| $OLV_{i,t}$ | | | 0.001 | | | | 0.001 |
| | | | (0.84) | | | | (0.80) |
| $QuoteInt_{i,t}$ | | | | 0.015** | | | -0.020*** |
| | | | | (2.26) | | | (-3.16) |
| $QT_{i,t}$ | | | | | 0.008** | | 0.008** |
| | | | | | (2.10) | | (2.37) |
| $MG_{i,t}$ | | | | | | 0.021*** | 0.032*** |
| | | | | | | (3.21) | (3.10) |
| Stock and Day FE | Yes | Yes | Yes | Yes | Yes | Yes | Yes |
| N obs. | 14,238 | 14,238 | 14,238 | 14,238 | 14,238 | 14,238 | 14,238 |
| $R^2$ | 3% | 0.1% | 0% | 0.4% | 0.2% | 0.7% | 3.3% |

| Panel B: $NASD_HFT_D_{i,t}$ | | | | | | | |
|---|---|---|---|---|---|---|---|
| | (i) | (ii) | (iii) | (iv) | (v) | (vi) | (vii) |
| $HFT_D_{i,t}$ | 0.068*** | | | | | | 0.083*** |
| | (3.70) | | | | | | (4.38) |
| $Flick_{i,t}$ | | -0.003*** | | | | | -0.003*** |
| | | (-3.27) | | | | | (-3.24) |
| $OLV_{i,t}$ | | | -0.000 | | | | -0.000 |
| | | | (-0.31) | | | | (-0.07) |
| $QuoteInt_{i,t}$ | | | | -0.002 | | | 0.030*** |
| | | | | (-0.42) | | | (2.60) |
| $QT_{i,t}$ | | | | | 0.016*** | | 0.019*** |
| | | | | | (3.34) | | (3.91) |
| $MG_{i,t}$ | | | | | | -0.006 | -0.040*** |
| | | | | | | (-1.16) | (-2.82) |
| Stock and Day FE | Yes | Yes | Yes | Yes | Yes | Yes | Yes |
| N obs. | 14,238 | 14,238 | 14,238 | 14,238 | 14,238 | 14,238 | 14,238 |
| $R^2$ | 0.8% | 0.1% | 0% | 0.5% | 0.5% | 0.3% | 1.4% |

| Panel C: $NASD_HFT_S_{i,t}$ | | | | | | | |
|---|---|---|---|---|---|---|---|
| | (i) | (ii) | (iii) | (iv) | (v) | (vi) | (vii) |
| $HFT_S_{i,t}$ | 0.246*** | | | | | | 0.239*** |
| | (38.28) | | | | | | (23.27) |
| $Flick_{i,t}$ | | -0.009*** | | | | | 0.001 |
| | | (-2.69) | | | | | (0.06) |
| $OLV_{i,t}$ | | | -0.004 | | | | 0.002* |
| | | | (-0.91) | | | | (1.68) |
| $QuoteInt_{i,t}$ | | | | 0.140*** | | | -0.006 |
| | | | | (14.02) | | | (-0.62) |
| $QT_{i,t}$ | | | | | 0.084*** | | 0.005 |
| | | | | | (5.41) | | (1.12) |
| $MG_{i,t}$ | | | | | | 0.144*** | 0.010 |
| | | | | | | (15.97) | (0.79) |
| Day FE | Yes | Yes | Yes | Yes | Yes | Yes | Yes |
| N obs. | 14,238 | 14,238 | 14,238 | 14,238 | 14,238 | 14,238 | 14,238 |
| $R^2$ | 74% | 0.3% | 0% | 51% | 15% | 53% | 74% |

| Panel D: $NASD_HFT_D_{i,t}$ | | | | | | | |
|---|---|---|---|---|---|---|---|
| | (i) | (ii) | (iii) | (iv) | (v) | (vi) | (vii) |
| $HFT_D_{i,t}$ | 0.423*** | | | | | | 0.383*** |
| | (23.23) | | | | | | (17.93) |
| $Flick_{i,t}$ | | -0.003 | | | | | -0.003 |
| | | (-0.41) | | | | | (-0.93) |
| $OLV_{i,t}$ | | | 0.000 | | | | 0.002** |
| | | | (0.24) | | | | (1.99) |
| $QuoteInt_{i,t}$ | | | | 0.091*** | | | 0.013 |
| | | | | (7.92) | | | (0.92) |
| $QT_{i,t}$ | | | | | 0.021** | | 0.014*** |
| | | | | | (2.04) | | (2.69) |
| $MG_{i,t}$ | | | | | | 0.092*** | 0.015 |
| | | | | | | (8.37) | (0.94) |
| Day FE | Yes | Yes | Yes | Yes | Yes | Yes | Yes |
| N obs. | 14,238 | 14,238 | 14,238 | 14,238 | 14,238 | 14,238 | 14,238 |
| $R^2$ | 50% | 0.1% | 0% | 23% | 1.0% | 24% | 54% |

**Table 4**
**Impact of exchange technological changes on HFT activity.**
This table examines how our ML-generated HFT measures respond to two technological changes: Nasdaq's reduced data dissemination latency and Amex's speed bump implementation. We estimate the following difference-in-difference models:

$$HFT_D_{i,t} = \alpha_i + \beta_t + \gamma_1 Post_{i,t} + \sum_{k=1}^{4} \delta_{i,t}^{k} C_{i,t}^{k} + \varepsilon_{i,t} \quad (4.1)$$
$$HFT_S_{i,t} = \alpha_i + \beta_t + \gamma_2 Post_{i,t} + \sum_{k=1}^{4} \delta_{i,t}^{k} C_{i,t}^{k} + \varepsilon_{i,t} \quad (4.2)$$
$$HFT_D_{i,t} = \alpha_i + \beta_t + \gamma_1 Post_{i,t} * Amex_{i,t} + \sum_{k=1}^{4} \delta_{i,t}^{k} C_{i,t}^{k} + \varepsilon_{i,t} \quad (4.3)$$
$$HFT_S_{i,t} = \alpha_i + \beta_t + \gamma_2 Post_{i,t} * Amex_{i,t} + \sum_{k=1}^{4} \delta_{i,t}^{k} C_{i,t}^{k} + \varepsilon_{i,t} \quad (4.4)$$

where $HFT_D_{i,t}$ and $HFT_S_{i,t}$ represent the ML-generated liquidity-demanding and -supplying HFT activities for stock $i$ on day $t$. $\alpha_i$ and $\beta_t$ capture stock and day fixed effects, respectively. For the Nasdaq upgrade analysis (Models 4.1 and 4.2), $Post_{i,t}$ equals 1 after October 10, 2011, for Nasdaq-listed stocks with tickers A-B, and after October 17, 2011, for other Nasdaq stocks. NYSE and Amex stocks serve as control groups in these models. For the Amex speed bump analysis (Models 4.3 and 4.4), $Post_{i,t}$ equals 1 after July 24, 2017, and $Amex_{i,t}$ equals 1 for Amex-listed stocks. NYSE and Nasdaq stocks serve as control groups in these models. Control variables ($C_{i,t}^{k}$) include daily volatility ($Volatility_{i,t}$, standard deviation of transaction-level returns), relative quoted spread ($Spread_{i,t}$, daily average of (ask-bid)/mid-quote for each transaction), inverse price ($InvPrice_{i,t}$), and dollar trading volume ($Volume_{i,t}$). The analysis uses 10-working day windows around implementation dates. Panel A reports results for the Nasdaq upgrade and Panel B for the Amex speed bump. Standard errors are double-clustered by stock and day, with $t$-statistics in brackets. *, **, and *** indicate significance at 10%, 5%, and 1%. $R^2$ values are within-$R^2$.

| | Panel A: Nasdaq upgrade | | Panel B: Amex speed bump | |
|---|---|---|---|---|
| | (i) $HFT_D_{i,t}$ | (ii) $HFT_S_{i,t}$ | (iii) $HFT_D_{i,t}$ | (iv) $HFT_S_{i,t}$ |
| $Post_{i,t}$ | 0.002** | 0.002** | | |
| | (2.12) | (2.10) | | |
| $Post_{i,t} * Amex_{i,t}$ | | | -0.005** | -0.007*** |
| | | | (-2.34) | (-3.31) |
| $Volatility_{i,t}$ | 0.013** | 0.000 | 0.001 | 0.001 |
| | (2.19) | (0.07) | (1.29) | (1.33) |
| $Spread_{i,t}$ | -0.066*** | -0.024*** | -0.015*** | -0.006*** |
| | (-12.58) | (-5.58) | (-10.96) | (-6.05) |
| $InvPrice_{i,t}$ | -0.151*** | 0.037 | -0.026 | -0.023* |
| | (-3.08) | (0.92) | (-1.57) | (-1.96) |
| $Volume_{i,t}$ | 0.001 | 0.020*** | 0.001** | 0.005*** |
| | (1.30) | (17.75) | (2.25) | (4.24) |
| Stock and Day FE | Yes | Yes | Yes | Yes |
| N obs. | 43,234 | 43,234 | 45,530 | 45,530 |
| $R^2$ | 5% | 11% | 1.3% | 3.5% |

**Table 5**
**HFT response to latency arbitrage opportunities.**
This table examines how our ML-generated HFT measures respond to latency arbitrage opportunities using the following OLS models:

$$HFT_D_{i,t} = \alpha_i + \beta_t + \gamma_1 NLAO_{i,t} + \sum_{k=1}^{4} \delta_{i,t}^{k} C_{i,t}^{k} + \varepsilon_{i,t}$$
$$HFT_S_{i,t} = \alpha_i + \beta_t + \gamma_2 NLAO_{i,t} + \sum_{k=1}^{4} \delta_{i,t}^{k} C_{i,t}^{k} + \varepsilon_{i,t}$$

where $HFT_D_{i,t}$ and $HFT_S_{i,t}$ represent our liquidity-demanding and -supplying HFT activity measures for stock $i$ and day $t$. $\alpha_i$ and $\beta_t$ capture stock and day fixed effects, respectively. $NLAO_{i,t}$ is the number of latency arbitrage opportunities, identified using the methodology detailed in Section 4.2. Control variables ($C_{i,t}^{k}$) include daily volatility ($Volatility_{i,t}$, standard deviation of transaction-level returns), relative quoted spread ($Spread_{i,t}$, daily average of (ask-bid)/mid-quote for each transaction), inverse price ($InvPrice_{i,t}$), and dollar trading volume ($Volume_{i,t}$). Columns (i) and (ii) present the results for $HFT_D_{i,t}$ and $HFT_S_{i,t}$, respectively. The sample consists of 120 randomly selected Nasdaq- and NYSE-listed firms. Standard errors are double-clustered by stock and day, with $t$-statistics in brackets. *, **, and *** indicate significance at 10%, 5%, and 1%. $R^2$ values are within-$R^2$.

| | (i)<br>$HFT_D_{i,t}$ | (ii)<br>$HFT_S_{i,t}$ |
|---|---|---|
| $NLAO_{i,t}$ | 0.018*** | -0.020** |
| | (3.78) | (-2.02) |
| $Volatility_{i,t}$ | -0.302*** | -0.353*** |
| | (-5.91) | (-4.50) |
| $Spread_{i,t}$ | -0.069*** | -0.033** |
| | (-4.66) | (-2.09) |
| $InvPrice_{i,t}$ | -0.390*** | 0.428*** |
| | (-6.04) | (7.98) |
| $Volume_{i,t}$ | -0.002*** | 0.003*** |
| | (-3.81) | (7.96) |
| Stock and Day FE | Yes | Yes |
| N obs. | 246,139 | 246,139 |
| $R^2$ | 17% | 12% |

**Table 6**
**Temporal stability of ML-generated HFT measures on Euronext.**
This table reports the results of regressions of actual Euronext HFT activity on ML-generated HFT predictions. The model is trained on 2009 Euronext data and used to predict actual HFT activity in 2010 and 2017. The regression specification is:

$$EU_HFT_S_{i,t} = \alpha_i + \beta_t + \gamma_1 HFT_S_{i,t} + \varepsilon_{i,t}$$
$$EU_HFT_D_{i,t} = \alpha_i + \beta_t + \gamma_1 HFT_D_{i,t} + \varepsilon_{i,t}$$

where $EU_HFT_D_{i,t}$ and $EU_HFT_S_{i,t}$ are actual share of liquidity-demanding or liquidity-supplying HFT volume from Bedofih and $HFT_D_{i,t}$ and $HFT_S_{i,t}$ are our ML-generated proxies (trained on 2009 Euronext data). Columns (1) and (2) report results for $HFT_S_{i,t}$ in 2010 and 2017, respectively. Columns (3) and (4) report results for $HFT_D_{i,t}$ in 2010 and 2017. The bottom of the table reports a $t$-test on the equality of the coefficients across the two years for each measure. The regressors are standardized, but the dependent variables are not. All specifications include stock and day fixed effects. Standard errors are double-clustered by stock and day, with $t$-statistics in brackets. *, **, and *** indicate significance at 10%, 5%, and 1%. $R^2$ values are within-$R^2$.

| | (1)<br>2010<br>$EU_HFT_S_{i,t}$ | (2)<br>2017<br>$EU_HFT_S_{i,t}$ | (3)<br>2010<br>$EU_HFT_D_{i,t}$ | (4)<br>2017<br>$EU_HFT_D_{i,t}$ |
|---|---|---|---|---|
| $HFT_S_{i,t}$ | 0.041*** | 0.045*** | | |
| | (3.67) | (3.63) | | |
| $HFT_D_{i,t}$ | | | 0.053*** | 0.077*** |
| | | | (5.54) | (7.58) |
| Stock and Day FE | Yes | Yes | Yes | Yes |
| N obs. | 67,193 | 73,089 | 67,193 | 73,089 |
| $R^2$ | 0.07% | 0.12% | 0.30% | 0.34% |
| $\Delta\gamma_1(2017 - 2010)$ | 0.004 | | 0.024* | |
| | (0.22) | | (1.76) | |

**Table 7**
**2SLS IV estimation results for JUMP.**
This table reports estimation results from the following models:

$$HFT_D_{i,m} = \beta_m + \gamma_1 LnPrice_{i,m-1} + \sum_{k=1}^{4} \delta_{i,m-1}^{k} C_{i,m-1}^{k} + \varepsilon_{i,m}$$

$$HFT_S_{i,m} = \beta_m + \gamma_1 LnPrice_{i,m-1} + \sum_{k=1}^{4} \delta_{i,m-1}^{k} C_{i,m-1}^{k} + \varepsilon_{i,m}$$

$$JUMP_{i,m} = \beta_m + \gamma_1 \widehat{HFT_D}_{i,m} + \sum_{k=1}^{4} \delta_{i,m-1}^{k} C_{i,m-1}^{k} + \varepsilon_{i,m}$$

$$JUMP_{i,m} = \beta_m + \gamma_1 \widehat{HFT_S}_{i,m} + \sum_{k=1}^{4} \delta_{i,m-1}^{k} C_{i,m-1}^{k} + \varepsilon_{i,m}$$

where $JUMP_{i,m}$ measures information acquisition for stock $i$ as the ratio of cumulative abnormal returns over [-1, 1] to cumulative abnormal returns over [-21, 1] around quarterly earnings announcements. $HFT_D_{i,m}$ and $HFT_S_{i,m}$ are our liquidity-demanding and liquidity-supplying HFT activities, measured as averages of daily values over [-21, -1] around earnings announcements. $LnPrice_{i,m-1}$ is the natural logarithm of the average daily prices over [-42, -22] around earnings announcements. Control variables ($C_{i,m-1}^{k}$) all measured as averages of daily values over [-42, -22] around earnings announcements, include volatility ($Volatility_{i,\mathrm{m}-1}$), relative effective spread ($ESpread_{i,\mathrm{m}-1}$), market value ($MValue_{i,\mathrm{m}-1}$, price times shares outstanding), and institutional order imbalance ($OIB20k_{i,\mathrm{m}-1}$, order imbalance of trades over \$20,000 from TAQ). Columns (i) and (ii) report the first-stage regressions of HFT demand and HFT supply on the log price instrument and control variables. Columns (iii) and (iv) report the second-stage IV regressions with the full set of controls. Columns (v) and (vi) report the IV regressions without the spread control. All variables are scaled by their root mean square about zero prior to estimation, without subtracting their means. The model is estimated at the monthly level on US-listed common stocks from 2010 to 2022, restricted to firm-quarters with quarterly earnings announcements. Month fixed effects are included. Standard errors are double-clustered by stock and month, with $t$-statistics in brackets. *, **, and *** indicate significance at 10%, 5%, and 1%. $R^2$ values are within-$R^2$.

| | (i)<br>$HFT_D_{i,m}$ | (ii)<br>$HFT_S_{i,m}$ | (iii)<br>$JUMP_{i,m}$ | (iv)<br>$JUMP_{i,m}$ | (v)<br>$JUMP_{i,m}$ | (vi)<br>$JUMP_{i,m}$ |
|---|---|---|---|---|---|---|
| $LnPrice_{i,m-1}$ | 0.221*** | -0.453*** | | | | |
| | (28.17) | (-22.97) | | | | |
| $\widehat{HFT_D}_{i,m}$ | | | 0.469*** | | 0.549*** | |
| | | | (6.97) | | (9.10) | |
| $\widehat{HFT_S}_{i,m}$ | | | | -0.229*** | | -0.306*** |
| | | | | (-6.89) | | (-8.54) |
| $Volatility_{i,m-1}$ | 0.023*** | 0.001 | -0.019*** | -0.001* | -0.046*** | -0.058*** |
| | (9.69) | (0.53) | (-3.71) | (-1.66) | (-5.34) | (-14.67) |
| $ESpread_{i,m-1}$ | -0.079*** | -0.030*** | -0.060*** | -0.104*** | | |
| | (-18.01) | (-6.20) | (-6.46) | (-16.00) | | |
| $MValue_{i,m-1}$ | 0.017*** | 0.139*** | -0.03*** | 0.01 | -0.033*** | 0.018*** |
| | (6.58) | (23.28) | (-5.02) | (1.45) | (-5.34) | (2.70) |
| $OIB20k_{i,m-1}$ | -0.419*** | -0.608*** | 0.235*** | -0.100*** | 0.242*** | -0.200*** |
| | (-80.82) | (-44.16) | (6.99) | (-5.79) | (7.20) | (-11.46) |
| Stock FE | No | No | No | No | No | No |
| Month FE | Yes | Yes | Yes | Yes | Yes | Yes |
| N obs. | 46,862 | 46,862 | 46,862 | 46,862 | 46,862 | 46,862 |
| $R^2$ | 79.28% | 54.56% | 0.12% | 2.65% | 0.01% | 1.54% |
| F-stat | 793.49 | 527.80 | | | | |

**Table 8**
**Comparing our HFT measures with Weller (2018) measures.**
This table analyzes the relationship between our ML-generated HFT measures and Weller's (2018) HFT proxies using the following OLS models:

$$CT_{i,m} = \alpha_i + \beta_m + \gamma_1 HFT_D_{i,m} + \gamma_2 HFT_S_{i,m} + \sum_{k=1}^{4} \delta_{i,m}^{k} C_{i,m}^{k} + \varepsilon_{i,t}$$

$$OLR_{i,m} = \alpha_i + \beta_m + \gamma_1 HFT_D_{i,m} + \gamma_2 HFT_S_{i,m} + \sum_{k=1}^{4} \delta_{i,m}^{k} C_{i,m}^{k} + \varepsilon_{i,t}$$

$$TO_{i,m} = \alpha_i + \beta_m + \gamma_1 HFT_D_{i,m} + \gamma_2 HFT_S_{i,m} + \sum_{k=1}^{4} \delta_{i,m}^{k} C_{i,m}^{k} + \varepsilon_{i,t}$$

The dependent variables are Weller's (2018) HFT proxies obtained from the MIDAS database: $CT_{i,m}$ (natural logarithm of cancel-to-trade ratio), $OLR_{i,m}$ (natural logarithm of odd-lot ratio), and $TO_{i,m}$ (natural logarithm of trade-to-order ratio), where each ratio is calculated as the average of daily values over [-21, -1] around earnings announcements. The key independent variables $HFT_D_{i,m}$ and $HFT_S_{i,m}$ are liquidity-demanding and liquidity-supplying HFT activities, measured as averages of daily values over [-21, -1] around earnings announcements. Models include stock ($\alpha_i$) and month ($\beta_m$) fixed effects, respectively. Control variables ($C_{i,m}^{k}$) all measured as averages of daily values over [-21, -1] around earnings announcements, include volatility ($Volatility_{i,m}$), relative effective spread ($ESpread_{i,m}$), market value ($MValue_{i,m}$, price times shares outstanding), and institutional order imbalance ($OIB20k_{i,m}$, order imbalance of trades over \$20,000 from TAQ). All variables are scaled by their root mean square about zero prior to estimation, without subtracting their means. The sample includes all U.S.-listed common stocks from 2012 to 2022. Standard errors are double-clustered by stock and quarter, with $t$-statistics in brackets. *, **, and *** indicate significance at 10%, 5%, and 1%. $R^2$ values are within-$R^2$.

| | (i)<br>$CT_{i,m}$ | (ii)<br>$OLR_{i,m}$ | (ii)<br>$TO_{i,m}$ |
|---|---|---|---|
| $HFT_D_{i,m}$ | 0.839*** | 2.714*** | -1.208*** |
| | (10.64) | (24.76) | (-15.71) |
| $HFT_S_{i,m}$ | -1.133*** | -2.343*** | 1.340*** |
| | (-12.01) | (-26.72) | (13.38) |
| $Volatility_{i,m}$ | 0.036 | -0.476*** | 0.492*** |
| | (0.74) | (-11.05) | (10.05) |
| $ESpread_{i,m}$ | -0.005 | 0.727*** | -0.190*** |
| | (-0.17) | (12.03) | (-5.76) |
| $MValue_{i,m}$ | 0.050*** | 0.120*** | -0.071*** |
| | (7.52) | (11.74) | (-8.63) |
| $OIB20k_{i,m}$ | 0.152*** | -0.111*** | 0.063* |
| | (5.24) | (-3.22) | (1.81) |
| Stock and Month FEs | Yes | Yes | Yes |
| N obs. | 43,091 | 43,091 | 43,091 |
| $R^2$ | 2% | 19% | 4% |

**Table 9.**
**2SLS IV estimation results for quoted spreads.**
This table reports estimation results from the following models:

$$HFT_D_{i,m} = \beta_m + \gamma_1 LnPrice_{i,m-1} + \sum_{k=1}^{4} \delta_{i,m-1}^{k} C_{i,m-1}^{k} + \varepsilon_{i,m}$$

$$HFT_S_{i,m} = \beta_m + \gamma_1 LnPrice_{i,m-1} + \sum_{k=1}^{4} \delta_{i,m-1}^{k} C_{i,m-1}^{k} + \varepsilon_{i,m}$$

$$Spread_{i,m} = \beta_m + \gamma_1 \widehat{HFT_D}_{i,m} + \sum_{k=1}^{4} \delta_{i,m-1}^{k} C_{i,m-1}^{k} + \varepsilon_{i,m}$$

$$Spread_{i,m} = \beta_m + \gamma_1 \widehat{HFT_S}_{i,m} + \sum_{k=1}^{4} \delta_{i,m-1}^{k} C_{i,m-1}^{k} + \varepsilon_{i,m}$$

where $Spread_{i,m}$ is the relative quoted spread, defined as the difference between the best ask and best bid divided by the midquote, averaged over month $m$ for stock $i$. $HFT_D_{i,m}$ and $HFT_S_{i,m}$ are our liquidity-demanding and liquidity-supplying HFT activities, measured as monthly averages of daily values for stock $i$ in month $m$. $LnPrice_{i,m-1}$ is the natural logarithm of the average daily prices in month $m-1$. Control variables ($C_{i,m-1}^{k}$) all measured as monthly averages of daily values for month $m-1$, include volatility ($Volatility_{i,m-1}$), relative effective spread ($ESpread_{i,m-1}$), market value ($MValue_{i,m-1}$, price times shares outstanding), and institutional order imbalance ($OIB20k_{i,m-1}$, order imbalance of trades over \$20,000 from TAQ). Columns (i) and (ii) report the first stage. Columns (iii) and (iv) report the second stage. All variables are scaled by their root mean square about zero prior to estimation, without subtracting their means. The model is estimated at the monthly level on US-listed common stocks from 2010 to 2023. Month fixed effects are included. Standard errors are double-clustered by stock and quarter, with $t$-statistics in brackets. *, **, and *** indicate significance at 10%, 5%, and 1%. $R^2$ values are within-$R^2$.

| | (i)<br>$HFT_D_{i,m}$ | (ii)<br>$HFT_S_{i,m}$ | (iii)<br>$Spread_{i,m}$ | (iv)<br>$Spread_{i,m}$ |
|---|---|---|---|---|
| $LnPrice_{i,m-1}$ | 0.205*** | -0.358*** | | |
| | (34.20) | (-26.56) | | |
| $\widehat{HFT_D}_{i,m}$ | | | 0.974*** | |
| | | | (16.21) | |
| $\widehat{HFT_S}_{i,m}$ | | | | -0.559*** |
| | | | | (-14.29) |
| $Volatility_{i,m-1}$ | 0.051*** | 0.003 | 0.067*** | 0.119*** |
| | (20.78) | (0.59) | (7.88) | (19.72) |
| $ESpread_{i,m-1}$ | -0.149*** | -0.054*** | 0.788*** | 0.612*** |
| | (-37.22) | (-7.49) | (41.47) | (60.59) |
| $MValue_{i,m-1}$ | 0.020*** | 0.136*** | -0.050*** | 0.045*** |
| | (8.98) | (25.83) | (-12.18) | (8.61) |
| $OIB20k_{i,m-1}$ | -0.449*** | -0.604*** | 0.647*** | -0.128*** |
| | (-113.99) | (-50.88) | (19.85) | (-7.10) |
| Stock FE | No | No | No | No |
| Month FE | Yes | Yes | Yes | Yes |
| N obs. | 553,248 | 553,248 | 553,248 | 553,248 |
| $R^2$ | 82.14% | 55.49% | 68.29% | 76.47% |
| F-stat | 1169.86 | 705.15 | | |

**Table 10**
**2SLS IV estimation results for quoted spreads during COVID-19.**
This table reports estimation results from the following models:

$$HFT_D_{i,d} = \beta_d + \gamma_1 LnPrice_{i,d-1} + \sum_{k=1}^{4} \delta_{i,d-1}^{k} C_{i,d-1}^{k} + \varepsilon_{i,d}$$

$$HFT_S_{i,d} = \beta_d + \gamma_1 LnPrice_{i,d-1} + \sum_{k=1}^{4} \delta_{i,d-1}^{k} C_{i,d-1}^{k} + \varepsilon_{i,d}$$

$$Spread_{i,d} = \beta_d + \gamma_1 \widehat{HFT_D}_{i,d} + \gamma_2 \widehat{HFT_D}_{i,d} \times Covid_d + \sum_{k=1}^{4} \delta_{i,d-1}^{k} C_{i,d-1}^{k} + \varepsilon_{i,d}$$

$$Spread_{i,d} = \beta_d + \gamma_1 \widehat{HFT_S}_{i,d} + \gamma_2 \widehat{HFT_S}_{i,d} \times Covid_d + \sum_{k=1}^{4} \delta_{i,d-1}^{k} C_{i,d-1}^{k} + \varepsilon_{i,d}$$

where $Spread_{i,d}$ is the relative quoted spread for stock $i$ on day $d$, defined as the difference between the best ask and best bid divided by the midquote. $HFT_D_{i,d}$ and $HFT_S_{i,d}$ are our liquidity-demanding and liquidity-supplying HFT activities for stock $i$ on day $d$. $LnPrice_{i,d-1}$ is the log of the closing price on day $d-1$. $Covid_d$ is a dummy that equals one for the period February 1 to March 31, 2020. Control variables ($C_{i,d-1}^{k}$), all lagged by one day, include volatility ($Volatility_{i,d-1}$), relative effective spread ($ESpread_{i,d-1}$), market value ($MValue_{i,d-1}$, price times shares outstanding), and institutional order imbalance ($OIB20k_{i,d-1}$, order imbalance of trades over $20,000 from TAQ). Columns (i) and (ii) report the first stage. Columns (iii) and (iv) report the baseline second stage without interactions. Columns (v) and (vi) report the second stage with COVID interactions. The standalone $Covid_d$ dummy is absorbed by the day fixed effects. All variables are scaled by their root mean square about zero prior to estimation, without subtracting their means. Indicator variables are not scaled. The sample covers three months before the COVID-19 recession, two months during, and three months after. Day fixed effects are included. Standard errors are double-clustered by stock and day, with $t$-statistics in brackets. *, **, and *** indicate significance at 10%, 5%, and 1%. $R^2$ values are within-$R^2$.

| | (i) $HFT_D_{i,d}$ | (ii) $HFT_S_{i,d}$ | (iii) $Spread_{i,d}$ | (iv) $Spread_{i,d}$ | (v) $Spread_{i,d}$ | (vi) $Spread_{i,d}$ |
|---|---|---|---|---|---|---|
| $LnPrice_{i,d-1}$ | 0.272*** | -0.359*** | | | | |
| | (33.86) | (-19.83) | | | | |
| $\widehat{HFT_D}_{i,d}$ | | | 0.695*** | | 0.696*** | |
| | | | (8.17) | | (8.19) | |
| $\widehat{HFT_S}_{i,d}$ | | | | -0.525*** | | -0.488*** |
| | | | | (-8.16) | | (-7.40) |
| $\widehat{HFT_D}_{i,d}$ * $Covid_d$ | | | | | -0.005 | |
| | | | | | (-0.14) | |
| $\widehat{HFT_S}_{i,d}$ * $Covid_d$ | | | | | | -0.138*** |
| | | | | | | (-3.70) |
| $Volatility_{i,d-1}$ | 0.006* | 0.007* | 0.172*** | 0.179*** | 0.172*** | 0.179*** |
| | (1.79) | (1.70) | (16.81) | (18.08) | (16.81) | (18.19) |
| $ESpread_{i,d-1}$ | -0.153*** | -0.153*** | 0.671*** | 0.483*** | 0.671*** | 0.483*** |
| | (-31.54) | (-22.10) | (23.84) | (33.82) | (23.92) | (33.65) |
| $MValue_{i,d-1}$ | 0.031*** | 0.169*** | -0.091*** | 0.02* | -0.090*** | 0.019* |
| | (10.83) | (25.89) | (-15.43) | (1.91) | (-15.40) | (1.85) |
| $OIB20k_{i,d-1}$ | -0.138*** | -0.215*** | 0.229*** | 0.02 | 0.229*** | 0.022* |
| | (-58.23) | (-40.34) | (16.04) | (1.59) | (16.03) | (1.68) |
| Stock FE | No | No | No | No | No | No |
| Day FE | Yes | Yes | Yes | Yes | Yes | Yes |
| N obs. | 445,943 | 445,943 | 445,943 | 445,943 | 445,943 | 445,943 |
| $R^2$ | 64.71% | 37.83% | 67.63% | 67.64% | 67.63% | 67.67% |
| F-stat | 1245.78 | 405.14 | | | | |

**Table 11.**
**2SLS IV estimation results for informed trading intensity.**
This table reports estimation results from the following models:

$$HFT_D_{i,m} = \beta_m + \gamma_1 LnPrice_{i,m-1} + \sum_{k=1}^{4} \delta_{i,m-1}^{k} C_{i,m-1}^{k} + \varepsilon_{i,m}$$

$$HFT_S_{i,m} = \beta_m + \gamma_1 LnPrice_{i,m-1} + \sum_{k=1}^{4} \delta_{i,m-1}^{k} C_{i,m-1}^{k} + \varepsilon_{i,m}$$

$$ITI_{i,m} = \beta_m + \gamma_1 \widehat{HFT_D}_{i,m} + \sum_{k=1}^{4} \delta_{i,m-1}^{k} C_{i,m-1}^{k} + \varepsilon_{i,m}$$

$$ITI_{i,m} = \beta_m + \gamma_1 \widehat{HFT_S}_{i,m} + \sum_{k=1}^{4} \delta_{i,m-1}^{k} C_{i,m-1}^{k} + \varepsilon_{i,m}$$

where $ITI_{i,m}$ is the informed trading intensity measure of Bogousslavsky et al. (2024) for stock $i$ in month $m$, computed as a monthly average of daily values. We use three versions of $ITI_{i,m}$: $ITI(13D)_{i,m}$, trained on the full 60-day Schedule 13D filing window and capturing overall informed trading; $ITI(Patient)_{i,m}$, trained on the first 40 days of the window and capturing informed trading that builds positions gradually; and $ITI(Impatient)_{i,m}$, trained on the last 20 days of the window and capturing more aggressive informed trading. $HFT_D_{i,m}$ and $HFT_S_{i,m}$ are our liquidity-demanding and liquidity-supplying HFT activities, measured as monthly averages of daily values for stock $i$ in month $m$. $LnPrice_{i,m-1}$ is the natural logarithm of the average daily prices in month $m-1$. Control variables ($C_{i,m-1}^{k}$) all measured as monthly averages of daily values for month $m-1$, include volatility ($Volatility_{i,m-1}$), relative effective spread ($ESpread_{i,m-1}$), market value ($MValue_{i,m-1}$, price times shares outstanding), and institutional order imbalance ($OIB20k_{i,m-1}$, order imbalance of trades over $20,000 from TAQ). Columns (i) and (ii) report the second-stage estimates for $ITI(13D)$, columns (iii) and (iv) for $ITI(Impatient)$, and columns (v) and (vi) for $ITI(Patient)$, respectively. All variables are scaled by their root mean square about zero prior to estimation, without subtracting their means. The model is estimated at the monthly level on US-listed common stocks from 2010 to 2023. Month fixed effects are included. Standard errors are double-clustered by stock and quarter, with $t$-statistics in brackets. *, **, and *** indicate significance at 10%, 5%, and 1%. $R^2$ values are within-$R^2$.

| | (i) $ITI(13D)_{i,m}$ | (ii) $ITI(13D)_{i,m}$ | (iii) $ITI(Impatient)_{i,m}$ | (iv) $ITI(Impatient)_{i,m}$ | (v) $ITI(Patient)_{i,m}$ | (vi) $ITI(Patient)_{i,m}$ |
|---|---|---|---|---|---|---|
| $\widehat{HFT_D}_{i,m}$ | -0.476*** | | -0.077*** | | -0.701*** | |
| | (-9.81) | | (-2.85) | | (-12.30) | |
| $\widehat{HFT_S}_{i,m}$ | | 0.184*** | | 0.030*** | | 0.271*** |
| | | (8.65) | | (2.82) | | (9.85) |
| $Volatility_{i,m-1}$ | 0.048*** | 0.024** | 0.019*** | 0.015*** | 0.054*** | 0.018*** |
| | (9.65) | (6.45) | (7.33) | (6.88) | (8.68) | (3.92) |
| $ESpread_{i,m-1}$ | -0.090*** | -0.001 | -0.032*** | -0.019*** | -0.118*** | 0.004 |
| | (-8.73) | (-1.29) | (-5.95) | (-5.55) | (-9.86) | (0.61) |
| $MValue_{i,m-1}$ | 0.003 | -0.027*** | -0.007*** | -0.012*** | 0.002 | -0.041*** |
| | (1.29) | (-8.42) | (-6.05) | (-6.95) | (1.20) | (-9.84) |
| $OIB20k_{i,m-1}$ | -0.182*** | 0.082*** | -0.051*** | -0.009* | -0.231*** | 0.159*** |
| | (-9.62) | (8.77) | (-4.61) | (-1.72) | (-10.39) | (13.02) |
| Stock FE | No | No | No | No | No | No |
| Month FE | Yes | Yes | Yes | Yes | Yes | Yes |
| N obs. | 364,875 | 364,875 | 364,875 | 364,875 | 364,875 | 364,875 |
| $R^2$ | 0.10% | 0.05% | 1.55% | 1.67% | 0.93% | 1.99% |

Online Appendix for **"Data-Driven Measures of High-Frequency Trading"**

August 2026

## Introduction

This online appendix provides supplementary results to the findings presented in Ibikunle et al. (2025). The content is as follows:

- Online Appendix A. Model Optimization and Machine Learning Comparisons
- Online Appendix B. Correlations Between HFT Measures and Market Quality Variables
- Online Appendix C. HFT Activity Around Earnings and M&A Announcements.
- Online Appendix D. Additional Tests on HFT and Information Acquisition
- Online Appendix E. Joint Estimation of HFT Demand and Supply Effects Using Heteroskedasticity-Based Instruments.
- Online Appendix F. HFT Activity and Quoted Spreads: Excluding Spread-Related Inputs from the HFT Measures
- Online Appendix G. Using Unscaled HFT Measures

# Online Appendix A. Model optimization and machine learning comparisons.

**Table OA.A.1**

**Parameter optimization results.**

The table lists the arithmetic mean and standard deviation for $R^2$ values across 10 iterations for different parameter combinations regarding the number of samples requires to split a tree node and the number of trees determining the ensemble size. Results are ranked by the Mean column.

| Rank | Mean | Std. | Split samples | Ensemble size |
|---|---|---|---|---|
| 1 | 0.814442 | 0.008260 | 5 | 640 |
| 2 | 0.813941 | 0.008360 | 5 | 320 |
| 3 | 0.813713 | 0.008455 | 5 | 160 |
| 4 | 0.812587 | 0.008609 | 5 | 80 |
| 5 | 0.810152 | 0.008016 | 5 | 40 |
| ... | ... | ... | ... | ... |
| 60 | 0.659040 | 0.027015 | 640 | 160 |
| 61 | 0.658566 | 0.022346 | 640 | 80 |
| 62 | 0.657760 | 0.022598 | 640 | 320 |
| 63 | 0.655796 | 0.023405 | 640 | 10 |
| 64 | 0.654791 | 0.027320 | 640 | 5 |

### OA.A.2. Comparison to simple machine learning methods.

We compare our ensemble methods against standard single-model alternatives to validate our approach. Since we predict continuous HFT outcomes rather than classify discrete categories, we test three benchmark models: LASSO, a linear regression that shrinks coefficients towards zero; support vector machines (SVM), which capture non-linear patterns through kernel functions; and neural networks with three hidden layers, which can learn complex relationships but require more data and computing power. For this comparison, we configure each model with reasonable default hyperparameters: LASSO uses alpha of 0.1 and tolerance of 0.0001, the SVM employs radial basis kernels, and the neural networks use rectified linear activation with mean absolute error optimisation. This allows us to benchmark whether tree ensembles outperform both simpler methods (LASSO, SVM) and more complex alternatives (deep learning) for HFT prediction.

Our dataset spans 29,880 stock-days, of which we drop 2,184 observations (7%) with missing dependent or independent variables. In these initial comparisons, we standardise each variable using z-score scaling to prevent variables with larger ranges from dominating predictions and to convert predictors to comparable units. We choose z-score over min-max scaling because HFT data contain extreme outliers that would distort min-max normalisation. Formally, for a dataset $D$,

$$z_{D_i} = \frac{D_i - \bar{D}}{\sigma(D)} \quad \text{(OA.A.1)},$$

We test two prediction approaches: multi-model (separate models for each target) versus multi-target (one model predicting both outcomes simultaneously). Multi-target models can capture relationships between our two dependent variables, potentially improving accuracy. However, LASSO and SVM require separate models by design, while neural networks can handle both approaches.

Online Appendix Table OA.A.2 reports mean $R^2$ values and standard deviations across 10 iterations for all methods under these default configurations. Extremely randomised trees deliver the highest mean $R^2$ (0.805) with a low standard deviation (0.035), outperforming both the simpler methods and the neural networks. While neural networks can theoretically approximate any function, they struggle to learn suitable parameters from our HFT dataset, a common challenge when financial data contain high noise relative to signal (Zhang et al. 2017).

Having identified extremely randomised trees as the best-performing architecture, we tune the model through a grid search with Monte Carlo cross-validation on the validation set; Online Appendix Table OA.A.1 reports the search space and selected values. The optimisation increases the number of component trees and produces finer node splits. Because node splits in tree-based models are unaffected by unscaled inputs, we no longer apply z-score scaling at this stage, so standard deviations are not directly comparable to the model comparisons above. The optimised model achieves an average $R^2$ of 0.825 with a standard deviation of 0.005 across multiple runs, a more accurate performance than the default configuration in Table OA.A.2. We therefore select extremely randomised trees as our primary method: it achieves superior prediction accuracy while remaining interpretable and computationally efficient.

**Table OA.A.2**
**Machine learning comparison.**
The table lists the arithmetic mean and standard deviation for $R^2$ values across 10 iterations for least absolute shrinkage and selection operator (LASSO), support vector regression (SVR), feed-forward artificial neural networks (ANN), random forests for multi-model (RF-MM) and multi-target (RF) setups, and extremely randomized trees for multi-model (ET-MM) and multi-target (ET) setups. Results are inversely ranked by the Mean column.

| Method | Mean | Std. |
|---|---|---|
| LASSO | 0.625 | 0.013 |
| SVR | 0.684 | 0.058 |
| ANN | 0.783 | 0.0229 |
| RF-MM | 0.784 | 0.055 |
| RF | 0.790 | 0.043 |
| ET-MM | 0.804 | 0.036 |
| ET | 0.805 | 0.035 |

## Online Appendix B. Correlations Between HFT Measures and Market Quality Variables.

**Table OA.B.1**
**Correlations between HFT measures and market quality variables.**
The table reports pairwise within-stock Pearson correlations between our ML-generated HFT measures ($HFT_D$ and $HFT_S$) and standard market quality variables. $Volatility$ is the daily standard deviation of transaction-level returns, $ESpread$ is the dollar value-weighted percentage effective spread for a given stock and day, $QSpread$ is the time-weighted percentage quoted spread during market hours for a given stock and day, $MValue$ is the market value, calculated as the end-of-day stock price multiplied by the number of shares outstanding, $Volume$ is the dollar volume, $Depth$ is the sum of the best ask size and the best bid size. The sample covers all US-listed common stocks from 2010 to 2023, at the daily level.

| | $HFT_D$ | $HFT_S$ | $Volatility$ | $ESpread$ | $QSpread$ | $MValue$ | $Volume$ | $Depth$ |
|---|---|---|---|---|---|---|---|---|
| $HFT_D$ | 1.000 | | | | | | | |
| $HFT_S$ | 0.497 | 1.000 | | | | | | |
| $Volatility$ | -0.183 | -0.109 | 1.000 | | | | | |
| $ESpread$ | -0.306 | -0.129 | 0.548 | 1.000 | | | | |
| $QSpread$ | -0.292 | -0.155 | 0.547 | 0.643 | 1.000 | | | |
| $MValue$ | 0.107 | 0.029 | -0.007 | -0.023 | -0.011 | 1.000 | | |
| $Volume$ | 0.204 | 0.248 | -0.015 | -0.032 | -0.039 | 0.471 | 1.000 | |
| $Depth$ | -0.019 | 0.356 | -0.006 | 0.072 | 0.047 | -0.121 | -0.063 | 1.000 |

## Online Appendix C. HFT Activity Around Earnings and M&A Announcements.

**Figure OA.C.1**
**HFT around earnings announcements.**
This figure illustrates the evolution of ML-generated HFT measures with their 95% confidence interval surrounding scheduled events, specifically earnings announcements. The event window spans 10 days before and after the announcement dates, which are sourced from the I/B/E/S database. The analysis encompasses all U.S. listed common stocks, with the sample period extending from 2010 to 2022.

Panel A: $HFT_S_{i,t}$ around earning announcements.

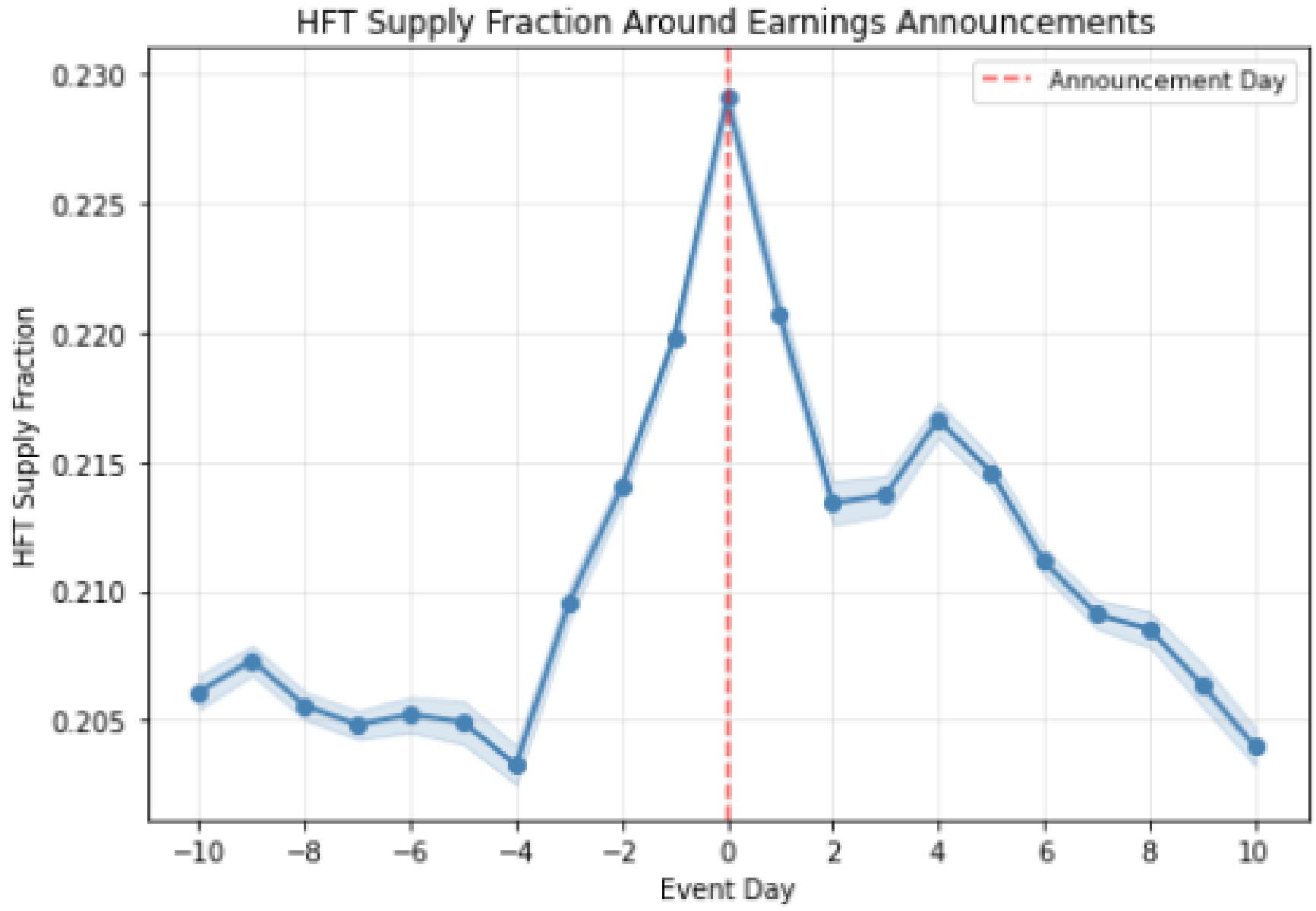

Panel B: $HFT_D_{i,t}$ around earning announcements.

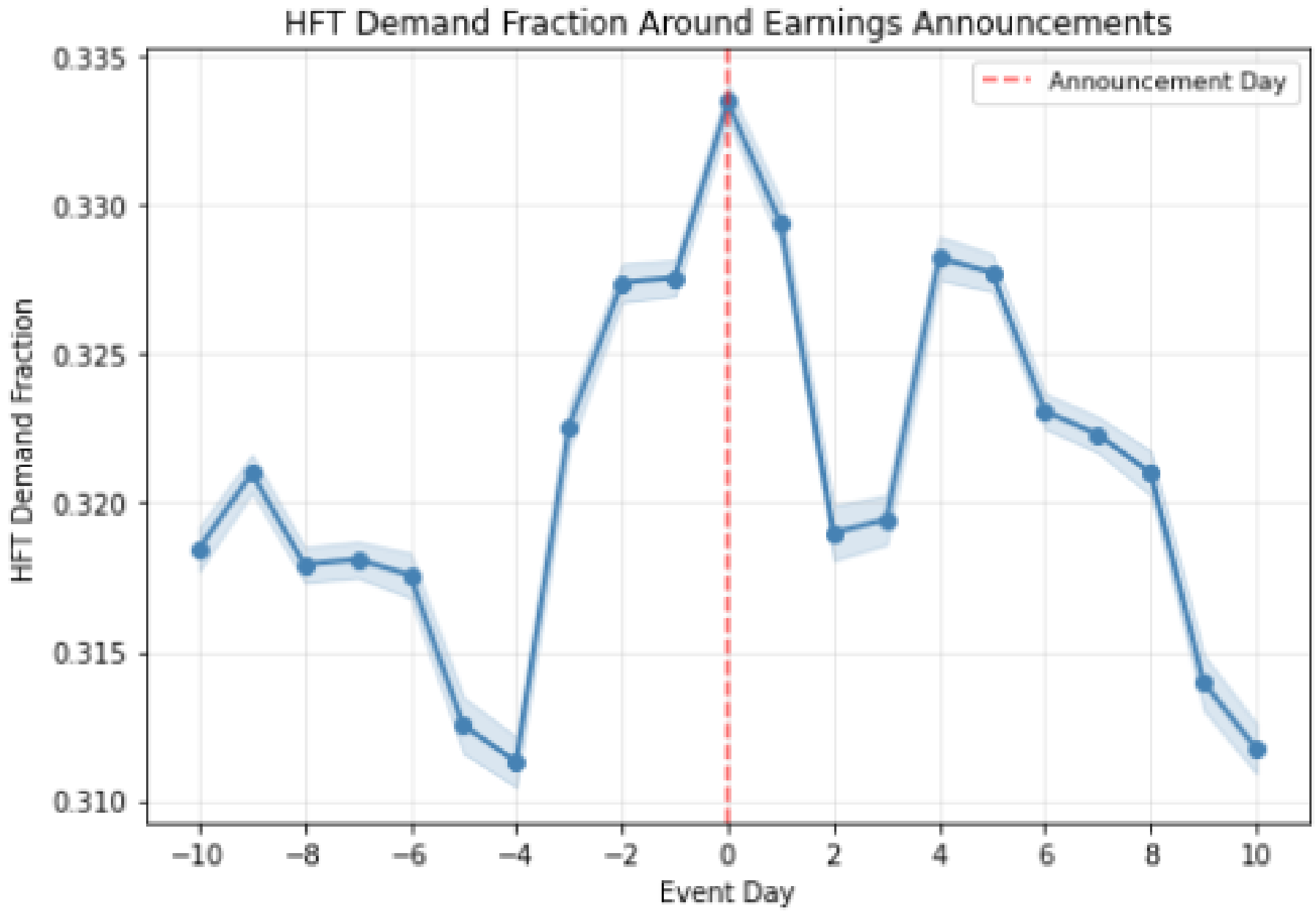

**Figure OA.C.2**
**HFT around M&A announcements.**
This figure illustrates the evolution of ML-generated HFT measures with their 95% confidence interval surrounding unscheduled events, specifically mergers and acquisitions (M&A) announcements. The event window spans 10 days before and after the announcement dates, which are sourced from the Thomson Reuters Securities Data Company (SDC) database. The analysis encompasses all U.S. listed common stocks, with the sample period extending from 2010 to 2022.

Panel A: $HFT_S_{i,t}$ around M&A announcements.

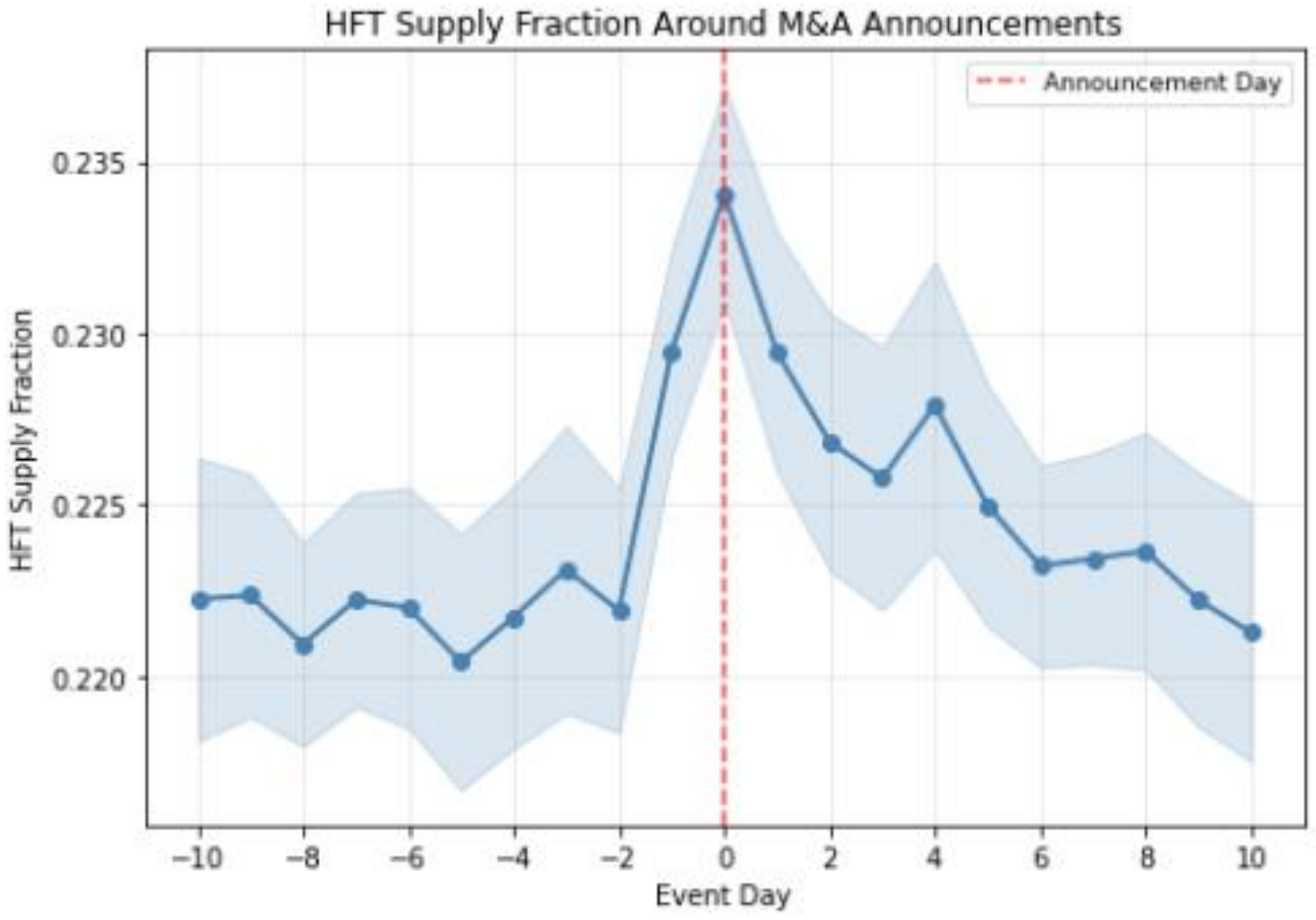

Panel B: $HFT_D_{i,t}$ around M&A announcements.

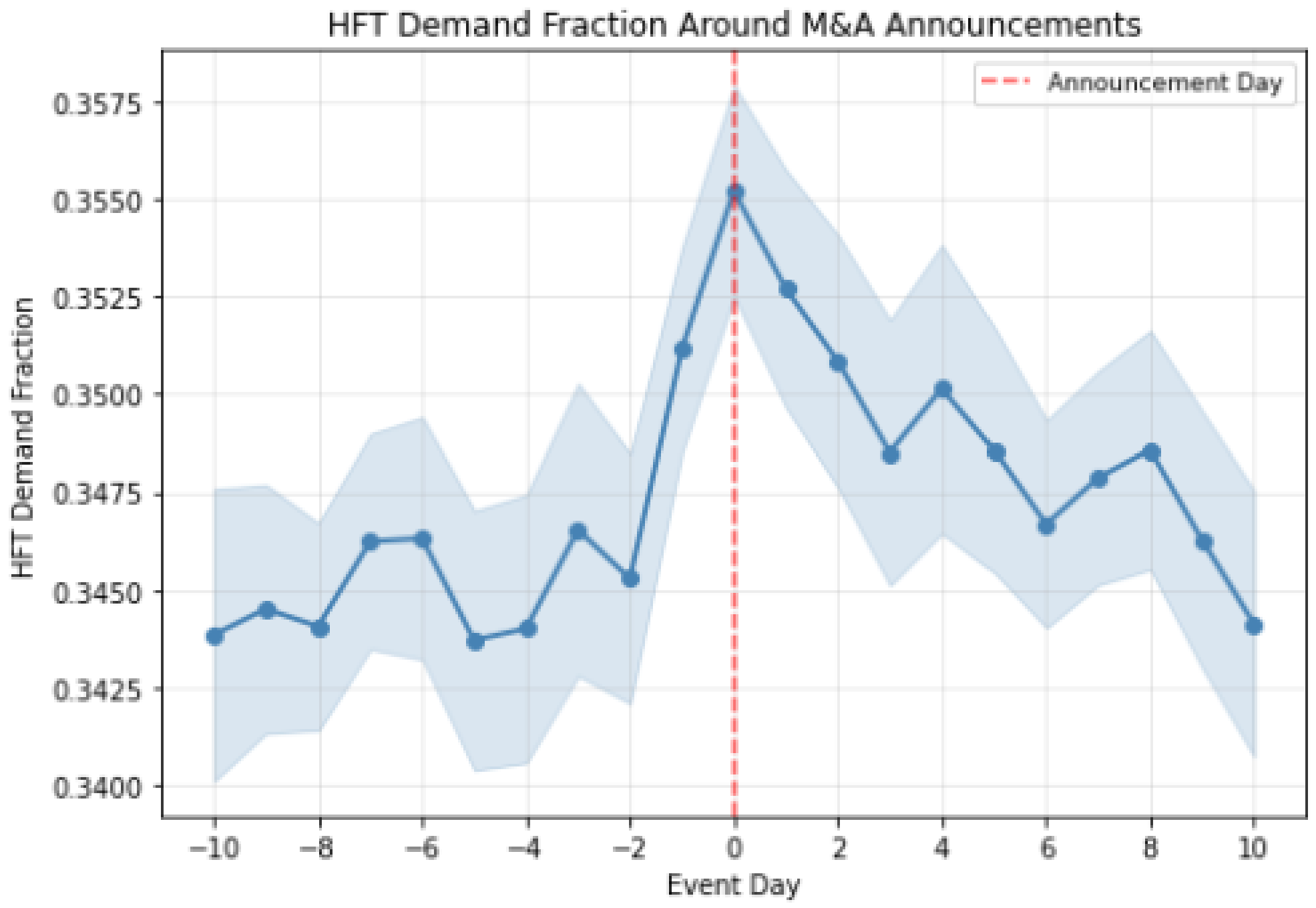

## Online Appendix D. Additional tests on HFT and information acquisition.

**Table OA.D.1**
**HFT activity and information acquisition using Nasdaq HFT data.**
This table replicates the analyses from Table 7 using Nasdaq's original HFT measures instead of our ML-generated proxies. $NASD_HFT_D_{i,m}$ and $NASD_HFT_S_{i,m}$ are Nasdaq's liquidity-demanding and -supplying HFT measures. The sample consists of 120 randomly selected stocks for which Nasdaq provided HFT data in 2009. All other specifications, including variable definitions, estimation methods, measurement periods, control variables, and fixed effects, remain identical to those in Table 7.

| | (i)<br>$JUMP_{i,m}$ | (ii)<br>$JUMP_{i,m}$ |
|---|---|---|
| $NASD\widehat{_HFT}_D_{i,m}$ | 4.317 | |
| | (0.14) | |
| $NASD\widehat{_HFT}_S_{i,m}$ | | 0.102 |
| | | (0.32) |
| Controls | As in Table 7 | As in Table 7 |
| Stock FE | No | No |
| Month FE | Yes | Yes |
| N obs. | 328 | 328 |
| $R^2$ | 0.0% | 1.5% |

**Table OA.D.2**
**HFT activity and information acquisition: analysis of 2010-2012 period.**
This table replicates the analyses from Table 7 using data from 2010 to 2012, a period immediately following our ML model's training sample (2009). All other specifications, including variable definitions, measurement periods, control variables, and fixed effects, remain identical to those in Table 7.

| | (i)<br>$JUMP_{i,m}$ | (ii)<br>$JUMP_{i,m}$ |
|---|---|---|
| $\widehat{HFT_D_{i,m}}$ | 0.449*** | |
| | (3.44) | |
| $\widehat{HFT_S_{i,m}}$ | | -0.196** |
| | | (-3.17) |
| Controls | As in Table 7 | As in Table 7 |
| Stock FE | No | No |
| Month FE | Yes | Yes |
| N obs. | 9,296 | 9,296 |
| $R^2$ | 0.2% | 2.7% |

## Online Appendix E. Joint Estimation of HFT Demand and Supply Effects Using Heteroskedasticity-Based Instruments.

**Table OA.E.1**

**Joint 2SLS estimation results using Lewbel (2012) instruments.**

This table reports second-stage 2SLS estimation results for our three main dependent variables: $JUMP_{i,m}$ in column (i), $Spread_{i,m}$ in column (ii), and $ITI(13D)_{i,m}$ in column (iii). $JUMP_{i,m}$ measures information acquisition for stock $i$ as the ratio of cumulative abnormal returns over [-1, 1] to cumulative abnormal returns over [-21, 1] around quarterly earnings announcements. $Spread_{i,m}$ is the relative quoted spread, calculated as the time-weighted average of the difference between the best ask and bid prices divided by the quote midpoint. $ITI(13D)_{i,m}$ is the informed trading intensity measure of Bogousslavsky et al. (2024), trained on the full 60-day Schedule 13D filing window. $HFT_D_{i,m}$ and $HFT_S_{i,m}$ are our liquidity-demanding and liquidity-supplying HFT activities. Unlike the main analysis, $HFT_D_{i,m}$ and $HFT_S_{i,m}$ enter each model jointly. The two endogenous variables are instrumented with three instruments: $LnPrice_{i,m-1}$ and two instruments constructed following Lewbel (2012). Each Lewbel instrument is the product of the demeaned $OIB20k_{i,m-1}$ and the residuals from one of the two first-stage regressions of $HFT_D_{i,m}$ and $HFT_S_{i,m}$ on $LnPrice_{i,m-1}$ and the control variables. $OIB20k_{i,m-1}$ is selected by regressing the squared first-stage residuals on the control variables and choosing the variable with the largest explanatory power for the error variance; the same variable is selected in all three specifications. Control variables include volatility ($Volatility_{i,m-1}$), relative effective spread ($ESpread_{i,m-1}$), market value ($MValue_{i,m-1}$, price times shares outstanding), and institutional order imbalance ($OIB20k_{i,m-1}$, order imbalance of trades over $20,000 from TAQ). For the $JUMP$ specification, the HFT measures are averages of daily values over [-21, -1], and $LnPrice_{i,m-1}$ and the control variables are averages of daily values over [-42, -22] around earnings announcements; for the $Spread$ and $ITI(13D)$ specifications, all variables are measured at the monthly level, with the instrument and controls lagged one month. All variables are scaled by their root mean square about zero prior to estimation, without subtracting their means. The model is estimated at the monthly level on US-listed common stocks from 2010 to 2023; the $JUMP$ specification is restricted to firm-quarters with quarterly earnings announcements from 2010 to 2022. Net effect reports the sum of the $HFT_D$ and $HFT_S$ coefficients, with the corresponding Wald test $t$-statistic in brackets. Month fixed effects are included. Standard errors are double-clustered by stock and month, with $t$-statistics in brackets. *, **, and *** indicate significance at 10%, 5%, and 1%. $R^2$ values are within-$R^2$.

| | (i)<br>$JUMP_{i,m}$ | (ii)<br>$Spread_{i,m}$ | (iii)<br>$ITI(13D)_{i,m}$ |
|---|---|---|---|
| $\widehat{HFT_D}_{i,m}$ | 0.350*** | 0.606*** | -0.167*** |
| | (5.48) | (13.28) | (-4.32) |
| $\widehat{HFT_S}_{i,m}$ | -0.057*** | -0.145*** | 0.075*** |
| | (-2.96) | (-12.83) | (8.22) |
| Controls | Yes | Yes | Yes |
| Stock FE | No | No | No |
| Month FE | Yes | Yes | Yes |
| N obs. | 46,862 | 553,248 | 368,475 |
| $R^2$ | 1.56% | 75.11% | 0.20% |
| $\widehat{HFT_D}_{i,m} + \widehat{HFT_S}_{i,m}$ | 0.293*** | 0.461*** | -0.092** |
| | (4.10) | (10.55) | (-2.29) |

## Online Appendix F. HFT Activity and Quoted Spreads: Excluding Spread-Related Inputs from the HFT Measures.

**Table OA.F.1**
**HFT activity and quoted spreads: Excluding spread-related inputs from the feature set.**
This table replicates the analyses in Table 9 using alternative HFT measures constructed from machine-learning models retrained after excluding four spread-related input variables: quoted spread, effective spread, realized spread, and price impact. The alternative liquidity-demanding and liquidity-supplying HFT measures are constructed using the same Nasdaq training sample and estimation procedure as the baseline measures and are then applied to all U.S.-listed common stocks from 2010 to 2023. Columns (i) and (ii) report the first-stage regressions, and columns (iii) and (iv) report the corresponding second-stage regressions for quoted spreads. The exclusion of these four variables applies only to the construction of the HFT measures; the lagged effective-spread control remains in the 2SLS regressions. All other variable definitions, measurement periods, instruments, controls, scaling procedures, fixed effects, and sample restrictions are identical to those in Table 9. Standard errors are double-clustered by stock and quarter, with $t$-statistics reported in brackets. *, **, and *** indicate significance at the 10%, 5%, and 1% levels. Reported $R^2$ values are within-$R^2$.

| | (i)<br>$HFT_D_{i,m}$ | (ii)<br>$HFT_S_{i,m}$ | (iii)<br>$Spread_{i,m}$ | (iv)<br>$Spread_{i,m}$ |
|---|---|---|---|---|
| $LnPrice_{i,m-1}$ | 0.183*** | -0.352*** | | |
| | (32.43) | (-25.78) | | |
| $\widehat{HFT_D}_{i,m}$ | | | 1.091*** | |
| | | | (15.59) | |
| $\widehat{HFT_S}_{i,m}$ | | | | -0.567*** |
| | | | | (-14.15) |
| Controls | Yes | Yes | Yes | Yes |
| Stock FE | No | No | No | No |
| Month FE | Yes | Yes | Yes | Yes |
| N obs. | 553,248 | 553,248 | 553,248 | 553,248 |
| $R^2$ | 83.24% | 52.47% | 66.53% | 75.76% |
| F-stat | 1051.83 | 664.58 | | |

## Online Appendix G. Using unscaled HFT measures.

**Table OA.G.1**

**Comparative analysis of HFT measures.**

This table evaluates our ML-generated unscaled HFT measures against alternative proxies using the following models:

$$NASD_U_HFT_D_{i,t} = \alpha_i + \beta_t + \gamma_1 U_HFT_D_{i,t} + \gamma_2 Flick_{i,t} + \gamma_3 OLV_{i,t} + \gamma_4 QuoteInt_{i,t} + \gamma_5 QT_{i,t} + \gamma_6 MG_{i,t} + \varepsilon_{i,t}$$
$$NASD_U_HFT_S_{i,t} = \alpha_i + \beta_t + \gamma_1 U_HFT_S_{i,t} + \gamma_2 Flick_{i,t} + \gamma_3 OLV_{i,t} + \gamma_4 QuoteInt_{i,t} + \gamma_5 QT_{i,t} + \gamma_6 MG_{i,t} + \varepsilon_{i,t}$$

where $NASD_U_HFT_D_{i,t}$ and $NASD_U_HFT_S_{i,t}$ are Nasdaq's unscaled liquidity-demanding and -supplying HFT measures, and $U_HFT_D_{i,t}$ and $U_HFT_S_{i,t}$ are our ML-generated unscaled HFT proxies (trained on January-June 2009 data), and alternative proxies from TAQ: quote volatility ($Flick_{i,t}$, average standard deviation of quote midpoints over 100 ms intervals), $OLV_{i,t}$ ($OLV_{i,t}$, sum of sub-100 share trades), quote intensity ($QuoteInt_{i,t}$, count of quote/depth changes), quote-to-trade ratio ($QT_{i,t}$, quoted shares/traded shares), and the number of messages ($MG_{i,t}$). The regressors are standardized, but the dependent variables are not. The analysis presents results for liquidity-supplying HFT in Panels A and C, while Panels B and D focus on liquidity-demanding HFT. Panels A and B incorporate both stock and day fixed effects, whereas Panels C and D employ only day fixed effect. The sample covers July-December 2009 for 120 randomly selected Nasdaq- and NYSE-listed firms with Nasdaq HFT data. Standard errors are double-clustered by stock and day, with $t$-statistics in brackets. *, **, and *** indicate significance at 10%, 5%, and 1%. $R^2$ values are within-$R^2$.

| Panel A: $NASD_U_HFT_S_{i,t}$ | | | | | | | |
|---|---|---|---|---|---|---|---|
| | (i) | (ii) | (iii) | (iv) | (v) | (vi) | (vii) |
| $U_HFT_S_{i,t}$ | 1.349*** | | | | | | 1.163*** |
| | (10.93) | | | | | | (9.14) |
| $Flick_{i,t}$ | | 0.002 | | | | | -0.002 |
| | | (0.78) | | | | | (-0.87) |
| $OLV_{i,t}$ | | | 0.012* | | | | 0.005*** |
| | | | (1.80) | | | | (2.59) |
| $QuoteInt_{i,t}$ | | | | 0.793*** | | | -0.073 |
| | | | | (5.89) | | | (-0.67) |
| $QT_{i,t}$ | | | | | -0.253*** | | -0.133** |
| | | | | | (-3.16) | | (-3.92) |
| $MG_{i,t}$ | | | | | | 0.940*** | 0.440*** |
| | | | | | | (5.88) | (2.88) |
| Stock and Day FE | Yes | Yes | Yes | Yes | Yes | Yes | Yes |
| N obs. | 14,238 | 14,238 | 14,238 | 14,238 | 14,238 | 14,238 | 14,238 |
| $R^2$ | 68% | 0.1% | 0.5% | 24% | 4.5% | 27% | 72% |

Panel B: $NASD_U_HFT_D_{i,t}$

| | (i) | (ii) | (iii) | (iv) | (v) | (vi) | (vii) |
|---|---|---|---|---|---|---|---|
| $U_HFT_D_{i,t}$ | 1.045*** | | | | | | 0.849*** |
| | (11.47) | | | | | | (9.75) |
| $Flick_{i,t}$ | | 0.002 | | | | | -0.005*** |
| | | (0.09) | | | | | (-3.14) |
| $OLV_{i,t}$ | | | 0.008 | | | | 0.004** |
| | | | (1.31) | | | | (2.45) |
| $QuoteInt_{i,t}$ | | | | 0.672*** | | | 0.090 |
| | | | | (6.12) | | | (1.15) |
| $QT_{i,t}$ | | | | | -0.179*** | | -0.100*** |
| | | | | | (-3.27) | | (-4.67) |
| $MG_{i,t}$ | | | | | | 0.788*** | 0.246** |
| | | | | | | (6.08) | (2.07) |
| Stock and Day FE | Yes | Yes | Yes | Yes | Yes | Yes | Yes |
| N obs. | 14,238 | 14,238 | 14,238 | 14,238 | 14,238 | 14,238 | 14,238 |
| $R^2$ | 64% | 0% | 0.5% | 31% | 4% | 33% | 69% |

Panel C: $NASD_U_HFT_S_{i,t}$

| | (i) | (ii) | (iii) | (iv) | (v) | (vi) | (vii) |
|---|---|---|---|---|---|---|---|
| $U_HFT_S_{i,t}$ | 1.552*** | | | | | | 1.534*** |
| | (31.38) | | | | | | (23.86) |
| $Flick_{i,t}$ | | -0.085*** | | | | | -0.001 |
| | | (-3.57) | | | | | (-0.49) |
| $OLV_{i,t}$ | | | -0.052 | | | | -0.006 |
| | | | (-1.21) | | | | (-0.79) |
| $QuoteInt_{i,t}$ | | | | 1.439*** | | | 0.082 |
| | | | | (6.04) | | | (0.50) |
| $QT_{i,t}$ | | | | | 1.054*** | | 0.022 |
| | | | | | (3.47) | | (0.71) |
| $MG_{i,t}$ | | | | | | 1.458*** | -0.070 |
| | | | | | | (6.01) | (-0.39) |
| Stock and Day FE | Yes | Yes | Yes | Yes | Yes | Yes | Yes |
| N obs. | 14,238 | 14,238 | 14,238 | 14,238 | 14,238 | 14,238 | 14,238 |
| $R^2$ | 90% | 0.3% | 0.1% | 60% | 25% | 61% | 95% |

| Panel D: $NASD_U_HFT_D_{i,t}$ | | | | | | | |
|---|---|---|---|---|---|---|---|
| | (i) | (ii) | (iii) | (iv) | (v) | (vi) | (vii) |
| $U_HFT_D_{i,t}$ | 1.167*** | | | | | | 1.146*** |
| | (27.10) | | | | | | (16.68) |
| $Flick_{i,t}$ | | -0.067*** | | | | | -0.002 |
| | | (-3.82) | | | | | (-1.62) |
| $OLV_{i,t}$ | | | -0.038 | | | | -0.005 |
| | | | (-1.06) | | | | (-0.28) |
| $QuoteInt_{i,t}$ | | | | 1.130*** | | | 0.245 |
| | | | | (7.68) | | | (1.55) |
| $QT_{i,t}$ | | | | | 0.710*** | | 0.022 |
| | | | | | (3.33) | | (0.95) |
| $MG_{i,t}$ | | | | | | 1.144*** | -0.226 |
| | | | | | | (7.63) | (-1.32) |
| Stock and Day FE | Yes | Yes | Yes | Yes | Yes | Yes | Yes |
| N obs. | 14,238 | 14,238 | 14,238 | 14,238 | 14,238 | 14,238 | 14,238 |
| $R^2$ | 92% | 0.4% | 0.1% | 70% | 22% | 71% | 94% |

**Table OA.G.2**
**Impact of exchange technological changes on HFT activity.**
This table examines how our ML-generated unscaled HFT measures respond to two technological changes: Nasdaq's reduced data dissemination latency and Amex's speed bump implementation. We estimate the following difference-in-difference models:

$$U_HFT_D_{i,t} = \alpha_i + \beta_t + \gamma_1 Post_{i,t} + \sum_{k=1}^{4} \delta_{i,t}^{k} C_{i,t}^{k} + \varepsilon_{i,t} \quad \text{(OA.G.1)}$$
$$U_HFT_S_{i,t} = \alpha_i + \beta_t + \gamma_2 Post_{i,t} + \sum_{k=1}^{4} \delta_{i,t}^{k} C_{i,t}^{k} + \varepsilon_{i,t} \quad \text{(OA.G.2)}$$
$$U_HFT_D_{i,t} = \alpha_i + \beta_t + \gamma_1 Post_{i,t} * Amex_{i,t} + \sum_{k=1}^{4} \delta_{i,t}^{k} C_{i,t}^{k} + \varepsilon_{i,t} \quad \text{(OA.G.3)}$$
$$U_HFT_S_{i,t} = \alpha_i + \beta_t + \gamma_2 Post_{i,t} * Amex_{i,t} + \sum_{k=1}^{4} \delta_{i,t}^{k} C_{i,t}^{k} + \varepsilon_{i,t} \quad \text{(OA.G.4)}$$

where $U_HFT_D_{i,t}$ and $U_HFT_S_{i,t}$ represent the ML-generated unscaled liquidity-demanding and -supplying HFT activities for stock $i$ on day $t$. $\alpha_i$ and $\beta_t$ capture stock and day fixed effects, respectively. For the Nasdaq upgrade analysis (Models OA.G.1 and OA.G.2), $Post_{i,t}$ equals 1 after October 10, 2011, for Nasdaq-listed stocks with tickers A-B, and after October 17, 2011, for other Nasdaq stocks. NYSE and Amex stocks serve as control groups in these models. For the Amex speed bump analysis (Models OA.G.3 and OA.G.4), $Post_{i,t}$ equals 1 after July 24, 2017, and $Amex_{i,t}$ equals 1 for Amex-listed stocks. NYSE and Nasdaq stocks serve as control groups in these models. Control variables ($C_{i,t}^{k}$) include daily volatility ($Volatility_{i,t}$, standard deviation of transaction-level returns), relative quoted spread ($Spread_{i,t}$, daily average of (ask-bid)/(0.5×(ask+bid) for each transaction)), inverse price ($InvPrice_{i,t}$), and dollar trading volume ($Volume_{i,t}$). The analysis uses 10-working day windows around implementation dates. Panel A reports results for the Nasdaq upgrade and Panel B for the Amex speed bump. Standard errors are double-clustered by stock and day, with $t$-statistics in brackets. *, **, and *** indicate significance at 10%, 5%, and 1%. $R^2$ values are within-$R^2$.

| | Panel A: Nasdaq upgrade | | Panel B: Amex speed bump | |
|---|---|---|---|---|
| | (i)<br>$U_HFT_D_{i,t}$ | (ii)<br>$U_HFT_S_{i,t}$ | (iii)<br>$U_HFT_D_{i,t}$ | (iv)<br>$U_HFT_S_{i,t}$ |
| $Post_{i,t}$ | 1.055***<br>(2.79) | 1.347**<br>(2.27) | | |
| $Post_{i,t} * Amex_{i,t}$ | | | -0.977**<br>(-2.27) | -0.696**<br>(-1.98) |
| Controls | Yes | Yes | Yes | Yes |
| Stock and Day FE | Yes | Yes | Yes | Yes |
| N obs. | 43,234 | 43,234 | 45,530 | 45,530 |
| $R^2$ | 29% | 18% | 59% | 49% |

**Table OA.G.3**
**HFT response to latency arbitrage opportunities.**
This table examines how our ML-generated unscaled HFT measures respond to latency arbitrage opportunities using the following OLS models:

$$U_HFT_D_{i,t} = \alpha_i + \beta_t + \gamma_1 NLAO_{i,t} + \sum_{k=1}^{4} \delta_{i,t}^{k} C_{i,t}^{k} + \varepsilon_{i,t}$$
$$U_HFT_S_{i,t} = \alpha_i + \beta_t + \gamma_2 NLAO_{i,t} + \sum_{k=1}^{4} \delta_{i,t}^{k} C_{i,t}^{k} + \varepsilon_{i,t}$$

where $U_HFT_D_{i,t}$ and $U_HFT_S_{i,t}$ represent the ML-generated unscaled liquidity-demanding and -supplying HFT activities for stock $i$ and day $t$. $\alpha_i$ and $\beta_t$ capture stock and day fixed effects, respectively. $NLAO_{i,t}$ is the number of latency arbitrage opportunities, identified using the methodology detailed in Section 4.2. Control variables ($C_{i,t}^{k}$) include daily volatility ($Volatility_{i,t}$, standard deviation of transaction-level returns), relative quoted spread ($Spread_{i,t}$, daily average of (ask-bid)/(0.5×(ask+bid) for each transaction), inverse price ($InvPrice_{i,t}$), and dollar trading volume ($Volume_{i,t}$). Columns (i) and (ii) present the results for $U_HFT_D_{i,t}$ and $U_HFT_S_{i,t}$, respectively. The sample consists of 120 randomly selected Nasdaq- and NYSE-listed firms. Standard errors are double-clustered by stock and day, with $t$-statistics in brackets. *, **, and *** indicate significance at 10%, 5%, and 1%. $R^2$ values are within-$R^2$.

| | (i)<br>$U_HFT_D_{i,t}$ | (ii)<br>$U_HFT_S_{i,t}$ |
|---|---|---|
| $NLAO_{i,t}$ | 66.266*** | -150.518** |
| | (3.21) | (-2.04) |
| Controls | Yes | Yes |
| Stock and Day FE | Yes | Yes |
| N obs. | 246,139 | 246,139 |
| $R^2$ | 39% | 38% |